\documentclass[pre,twocolumn,superscriptaddress, aps]{revtex4-2}
\usepackage[utf8]{inputenc}

\usepackage[separate-uncertainty = true,multi-part-units=single]{siunitx}
\usepackage{changes}
\usepackage{graphicx}
\usepackage{amsmath}
\usepackage{amssymb}
\usepackage[hidelinks]{hyperref}

\usepackage[version=4]{mhchem}
\usepackage{chemformula}
\setchemformula{kroeger-vink}
\DeclareSIUnit\angstrom{\text{Å}}

\presetkeys%
    {todonotes}%
    {inline,backgroundcolor=white}{}
\makeatletter

\makeatletter
\renewcommand{\todo}[2][]{\@bsphack\@todo[#1]{\textcolor{black!70}{#2}}\@esphack\ignorespaces}
\makeatother

\begin{document}

\title{Mechanism and kinetics of sodium diffusion in Na-feldspar from neural network based atomistic simulations}

\author{Alexander Gorfer}
\affiliation{Faculty of Physics, University of Vienna, Boltzmanngasse 5, 1090, Vienna, Austria}
\affiliation{Department of Lithospheric Research, University of Vienna, Josef-Holaubuek-Platz 2, 1090, Vienna, Austria}
\affiliation{Vienna Doctoral School in Physics, University of Vienna, Boltzmanngasse 5, 1090, Vienna, Austria}
\author{Rainer Abart}
\affiliation{Department of Lithospheric Research, University of Vienna, Josef-Holaubuek-Platz 2, 1090, Vienna, Austria}
\author{Christoph Dellago}
\affiliation{Faculty of Physics, University of Vienna, Boltzmanngasse 5, 1090, Vienna, Austria}

\begin{abstract}
Alkali diffusion is a first-order control for microstructure and compositional evolution of feldspar during cooling from high temperatures of primary magmatic or metamorphic crystallization, and knowledge of the respective diffusion coefficients is crucial for reconstructing thermal histories. Our understanding of alkali diffusion in feldspar is, however, hindered by an insufficient grasp of the underlying diffusion mechanisms. We performed molecular dynamics simulations of sodium feldspar (Albite) containing different point defects using a recently developed neural network potential. A high degree of agreement between the sodium self-diffusion coefficients obtained from model simulations and those determined experimentally in earlier studies  motivated a detailed investigation into the interstitial and vacancy mechanisms, corresponding jump rates, correlation factors and anisotropy. We identified a dumbbell shaped double occupancy of an alkali site as an important point defect and a correlation effect originating from the orientation of the dumbbell as a possible cause for the \(\perp\!\!(001) > \, \perp\!\!(010)\) diffusion anisotropy, which has been reported in a slew of feldspar cation diffusion experiments.
\end{abstract}
\maketitle

\section{Introduction}
Feldspars are abundant rock-forming minerals, which may be described as an essentially ternary solid solution between sodium (Albite: \ch{NaAlSi3O8}), potassium  (K-Feldspar: \ch{KAlSi3O8}) and calcium (Anorthite: \ch{CaAl2Si2O8}) endmembers. The crystal structure of alkali feldspar, the binary solid-solution between Na-feldspar and K-Feldspar, is comprised of a 3D framework of corner-sharing [SiO$_4$] and [AlO$_4$] tetrahedra, and the alkali cations occupy large, irregularly coordinated sites within the framework. The alkali cations are substantially more mobile than Al and Si on the tetrahedral sites. Diffusion of alkali cations in alkali feldspar is of pivotal importance for the adjustment of the alkali cation partitioning with other phases and thus for the reconstruction of formation conditions from phase equilibria~\cite{spear1994metamorphic}. Furthermore, alkali cation diffusion controls the rates of exsolution of originally homogeneous alkali feldspar of intermediate composition into perthite, a lamellar intergrowth of  Na-rich and  K-rich alkali feldspar. From the widths and chemical compositions of the exsolution lamellae the rock's thermal history can be inferred~\cite{Abart2009,Petrishcheva2009Exsolution,PETRISHCHEVA20125481}). Similarly, alkali diffusion in alkali feldspar and grain internal microstructures from exsolution are highly relevant for the retention of radiogenic isotopes and thus for application of the \ch{Rb-Sr}, the \ch{^{40}Ar-^{39}Ar}~\cite{WARTHO1999141}, the \ch{K-Ar} and the \ch{K-Ca} geochronological systems. %

The mechanism of cation diffusion in feldspar has been a topic of controversy for over half a century. Comprehensive summaries of this topic were given by Smith~\cite{smith1974feldspar}, Yund~\cite{Yund1983}, Smith and Brown~\cite{JosephV.Smith1988} and most recently by Cherniak~\cite{Cherniak2010}. Diffusion experiments are complicated by the range of possible compositions and states of Al-Si disorder in the tetrahedral framework. Moreover, natural specimens need to be used as starting materials, because synthesis of sufficiently large crystals with compositions and states of Al-Si disorder corresponding to those of natural crystals is not feasible. In addition, the low symmetry which can vary between mono- and triclinic and diffusion ansiotropy further complicate experimental determination of alkali cation diffusivities.  A complete diffusivity tensor has only been determined by Petrishcheva and coworkers \cite{Petrishcheva2014} for Na-K interdiffusion at \SI{850}{\celsius}, where pronounced anisotropy with diffusivity in the \([101]\) direction up to about 10 times higher, depending on composition, than in the \(\perp\!\!(10\overline{1})\) direction was observed. This is in contrast to \ch{Na^+} and \ch{K^+} self-diffusion, which has been found to be close to isotropic. Tracer-diffusion in \(\perp\!\!(010)\) and \(\perp\!\!(001)\) is almost identical in Ca-rich and labradoritic \cite{Behrens1990} feldspars. The ionic conductivity in K-rich alkali feldspar~\cite{ElMaanaoui2016} however supports anisotropy as the \(\perp\!\!(001)\) direction is distinctly faster than \(\perp\!\!(010)\). %
Self-diffusion coefficients along additional directions, particularly in Na-rich feldspar, are needed for a better understanding of diffusion anisotropy. One application are diffuse interface models of exsolution in alkali feldspar \cite{Abart2009} which did not yet include effects of diffusion anisotropy.

Different mechanisms for alkali cation transport have been considered over the years, all related to point defects \cite{ElMaanaoui2016, wilangowski2015_proceeding, giletti_alkali_1997, Behrens1990, Petrovic1974, foland_alkali_1974}. In alkali feldspar, the most important type of point defect is likely the Frenkel pair, which consists of a \ch{Na^+} or \ch{K^+} interstitial and a vacancy on an alkali site. The Schottky defect, made up of two vacancies of opposite sign, is expected to be less favourable due to the strong bonds between Al or Si and O in the tetrahedral framework. For the Frenkel pair, a likely position of the interstitial was presumed by Petrovic~\cite{Petrovic1974} to lie at the \((0,0,\tfrac{1}{2})\) position of the conventional unit cell (as shown in Fig.~\ref{fig:Cell_and_detection}). Based on DFT calculations, this configuration was recently confirmed by the present authors \cite{gorfer2024structure} to be stable in the case of sodium interstitials \ch{Na^+_I} in Na-feldspar. However, an energetically more favourable configuration was also found, where an alkali site is occupied by two sodium cations forming a dumbbell-type interstitial. 

So far, only the vacancy mechanism of cation transport has been studied using atomistic simulations by Jones et al.~\cite{Jones2004} or using Monte Carlo simulations by Wilangowski and Stolwijk~\cite{WilangowskiandStolwijk2015}. %
However, at least in the case of Ca-rich \cite{Behrens1990} and K-rich feldspars \cite{wilangowski2015_proceeding}, measurements of Haven ratios support the notion that transport of \ch{Na^+} is dominated by interstitials rather than by vacancies. Concurrently, theoretical models to interpret Haven ratios were based on the interstitialcy mechanism \cite{ElMaanaoui2016} in which the charge moves twice as far for a jump of the tracer, contrary to diffusion via the dumbbell mediated cation transport mechanism presented in this work.

In summary, there is currently little understanding of the migration mechanism of alkali cations in alkali feldspar on the atomistic level.

In the present work, we aim at revealing the atomistic mechanisms of self-diffusion of \ch{Na^+} through vacancy and interstitial migration in Na-feldspar. To this end, we use molecular dynamics (MD) to simulate systems of Na-feldspar containing a Na-vacancy, \ch{V^-_{Na}}, or a Na-interstitial, \ch{Na^+_{I}} defect. The sodium self-diffusion coefficient \(D_{\ch{Na}}\) that we consider in this study is therefore based only an these two defects \cite{Seeger1968},
\begin{equation}\label{Eq:Diffusion_sd}
    \begin{split}
        D_{\ch{Na}} &= D_{\mathrm{I}} f_\mathrm{I} C_{\mathrm{FP}} + D_{\mathrm{V}} f_\mathrm{V} C_{\mathrm{FP}},\\
    \end{split}
\end{equation}
where \(D_{\mathrm{I}}\) and \(D_{\mathrm{V}}\) are the diffusion coefficients of the interstital and the vacancy, respectively,  \(f_{\mathrm{I}}\) and \(f_{\mathrm{V}}\) are their respective correlation factors and \(C_{\mathrm{FP}} = C_{\mathrm{I}}=C_{\mathrm{V}}\) is the concentration of Frenkel pairs, which is necessarily equal to the concentration of interstitials, \(C_{\mathrm{I}}\), and vacancies, \(C_{\mathrm{V}}\). The correlation factors \(f_{\mathrm{I}}\) and \(f_{\mathrm{V}}\) in the above equation are needed to take into account that consecutive jumps of a tracer Na-cation are correlated due to successive interactions with interstitials and vacancies and do not follow a random walk. 
Equation (\ref{Eq:Diffusion_sd}) neglects different charge states of the defects (of which the neutral one has been shown in \cite{gorfer2024structure} to be unfavourable), defect clusters and  contributions of other defects such as Schottky defects. The concerted exchange mechanism was not observed in our simulations.

The interatomic potential driving the MD simulation is of crucial importance to correctly model the migration mechanism and defect concentrations and enable comparison with available experimental data. In this work we use a neural network potential (NNP) that we developed previously for Na-feldspar~\cite{gorfer2024structure}. This potential  reproduces results of DFT calculations but at a much lower computational cost, allowing for the long simulation timescales necessary to study cation diffusion. %
Since the concentration of Frenkel defects \(C_{\mathrm{FP}}\) was already determined in~\cite{gorfer2024structure}, in this study we focus on the self-diffusion constant of Na cations, \(D_{\ch{Na}}\), and on the diffusion constants of interstitials and vacancies, \(D_{\mathrm{I}}\) and \(D_{\mathrm{V}}\), as well as their correlation factors \(f_{\mathrm{I}}\) and \(f_{\mathrm{V}}\). %

The remainder of the article is structured as follows: In Section~\ref{sec:methods} we describe the computational methods (A), including an algorithm to localize the point defects (B) and introduce a lattice description of diffusion (C). Results are discussed in Section~\ref{sec:results}, which focuses on jump rates (A), self diffusion and correlation (B) and ends with a characterization of the interstitial (C). Conclusions are provided in Section~\ref{sec:conclusions}.

\section{Methods}\label{sec:methods}

\subsection{Simulation Setup and Definitions}\label{sec:simulation_definitions}

In this work we use a neural network potential committee (NNP) that was trained previously for Na-feldspar with its point defects and incorporates electrostatic corrections \cite{gorfer2024structure}. This NNP was trained with reference data obtained in the temperature range 900 - 1400 \si{\kelvin} and at atmospheric pressure, which is used throughout this work. Simulations were performed with the LAMMPS suite~\cite{Thompson2022}, patched with the n2p2p extension~\cite{Singraber2019} to use the NNP.

We first determined the simulation box parameters at the target temperatures, 900 to \SI{1400}{\kelvin} in steps of \SI{100}{\kelvin} in the case of \ch{Na^+_{I}} and 1000 to \SI{1400}{\kelvin} in case of \ch{V^-_{Na}}, and atmospheric pressure. For this purpose, we initialized a pristine Albite system made up of \begin{small}\(2\!\times\!2\!\times\!3\)\end{small} conventional unit cells (illustrated in Fig.~\ref{fig:Cell_and_detection}~a) containing 624 atoms in total using the box parameters described in~\cite{gorfer2024structure}. Either a \ch{V^-_{Na}} or an \ch{Na^+_{I}} at a dumbbell site was added. Then, an MD simulation in the NPT ensemble was performed for \SI{10}{\nano\second} using a Nose-Hoover barostat with a damping parameter of \SI{1}{\pico\second} and a Langevin thermostat with a damping parameter of \SI{0.1}{\pico\second} for each target temperature. The averages of the simulation box dimensions were calculated over the latter half of this trajectory and new defect-containing systems were initialized with these equilibrated box dimensions.

The box-equilibrated systems were equilibrated in the NVT ensemble for \SI{1}{\nano\second} using a Langevin thermostat. The thermostat was then switched off and configurations were produced in the NVE ensemble for at least \SI{21}{\nano\second}. The NVT equilibration and NVE production were repeated 5 times with different random seeds in the thermostat. The time step was \SI{1}{\femto\second} and configurations containing positions of all atoms were saved every \SI{25}{\femto\second}.

This trajectory with a length of \(N_t\) timesteps, containing \(N_{\ch{Na}}\) equal to \(48 \pm 1\) sodium ions and a single \ch{V^-_{Na}} or a \ch{Na^+_{I}} allows us to characterize the mobility of sodium using the sum of time lagged squared displacements (per interstitial):
\begin{equation}
    \mathrm{ssd}^{(\mathrm{I})}_{\ch{Na}}(m) = \frac{1}{N_t - m} \sum_{i=1}^{N_{\ch{Na}}} \sum_{k=0}^{N_t - m - 1}\!\!\! \left[\mathbf{r}_\mathrm{i}(k + m) - \mathbf{r}_\mathrm{i}(k)\right]^2\!\!\!,
\end{equation}
where \(\mathbf{r}_\mathrm{i}(k)\) is the position of the \(i\)th sodium ion at timestep \(k\). Note that we average over the \(N_t - m\) possible time differences but not over the particles as the non-ballistic motion of all sodium atoms is completely attributed to the defect such that \(\mathrm{ssd}^{(\mathrm{I})}_{\ch{Na}}(m)\) for long \(m\) is independent of system size.

The self-diffusivity per interstitial or vacancy can then be calculated by fitting \(\mathrm{ssd}^{(\mathrm{I})}_{\ch{Na}}\) to a linear model using the Einstein-relation
\begin{equation}\label{Eq:SelfD^I}
    D^{(\mathrm{I})}_{\ch{Na}} = \frac{1}{2d}\lim_{m \rightarrow \infty}\frac{\mathrm{d}}{\mathrm{d}t}\mathrm{ssd}^{(\mathrm{I})}_{\ch{Na}}(m),
\end{equation}
where \(d\) is the dimensionality of \(\mathbf{r}\). The same can be done for the vacancy, yielding \(\mathrm{ssd}^{(\mathrm{V})}_{\ch{Na}}\) and \(D^{(\mathrm{V})}_{\ch{Na}}\).

The diffusivity of the point defects is calculated in a similar manner. If \(\mathbf{r}_\mathrm{I}\) is the position of the interstitial (defined in the following section), then 
\begin{equation}
    \mathrm{sd}_{\mathrm{I}}(m) = \frac{1}{N_t - m} \sum_{k=0}^{N_t - m - 1} \left[\mathbf{r}_\mathrm{I}(k + m) - \mathbf{r}_\mathrm{I}(k)\right]^2,
\end{equation}
yielding the diffusion coefficient of an interstitial
\begin{equation}\label{Eq:D_I}
    D_\mathrm{I} = \frac{1}{2d}\lim_{m \rightarrow \infty}\frac{\mathrm{d}}{\mathrm{d}t}\mathrm{sd}_{\mathrm{I}}(m)
\end{equation}
and likewise for the vacancy.

The self-diffusivity per intersitial \(D^{(\mathrm{I})}_{\ch{Na}}\) and the interstitial diffusivity \(D_\mathrm{I}\) are related through the correlation factor:
\begin{equation}\label{Eq:Def_Correlationfactors}
    D^{(\mathrm{I})}_{\ch{Na}} = f_\mathrm{I} D_\mathrm{I}, ~~~~~~~ D^{(\mathrm{V})}_{\ch{Na}} = f_\mathrm{V} D_\mathrm{V}.
\end{equation}
These correlation factors %
can only be equated to a Bardeen-Herring correlation factor, which may be defined as the ratio of the diffusivity of tagged atoms and a diffusivity coming out of an uncorrelated jump sequence \cite{mehrer_diffusion_2007}, when the defect itself performs an uncorrelated walk. While this is very much expected in the case of self-diffusion, insofar that the difference is usually not commented upon (e.g.~\cite{Posselt2008}), a quite unusual result in the Monte Carlo study of Wilangowski and Stolwijk~\cite{WilangowskiandStolwijk2015} suggests that vacancies themselves are slightly correlated in the \(\perp\!\!(001)\) direction of alkali feldspar. This effect would however require further study and is not considered here.

Experimental methods used in the present literature such as tracer diffusion \cite{wilangowski2015_proceeding} or isotope absorption \cite{Kasper1975} do not measure any of the above coefficients, but rather the total self-diffusion coefficient per molar unit given by
\begin{equation}\label{eq:na_selfdiffusion}
    D_{\ch{Na}} = C_{\mathrm{FP}} \left(D^{(\mathrm{I})}_{\ch{Na}} + D^{(\mathrm{V})}_{\ch{Na}}\right),
\end{equation}
where \(C_{\mathrm{FP}}\) is the concentration of Frenkel pairs from \cite{gorfer2024structure}.

\subsection{Localization of Point Defects}

\begin{figure}[h]
    \centering
    \makebox[\columnwidth][c]{\includegraphics[width=1\columnwidth]{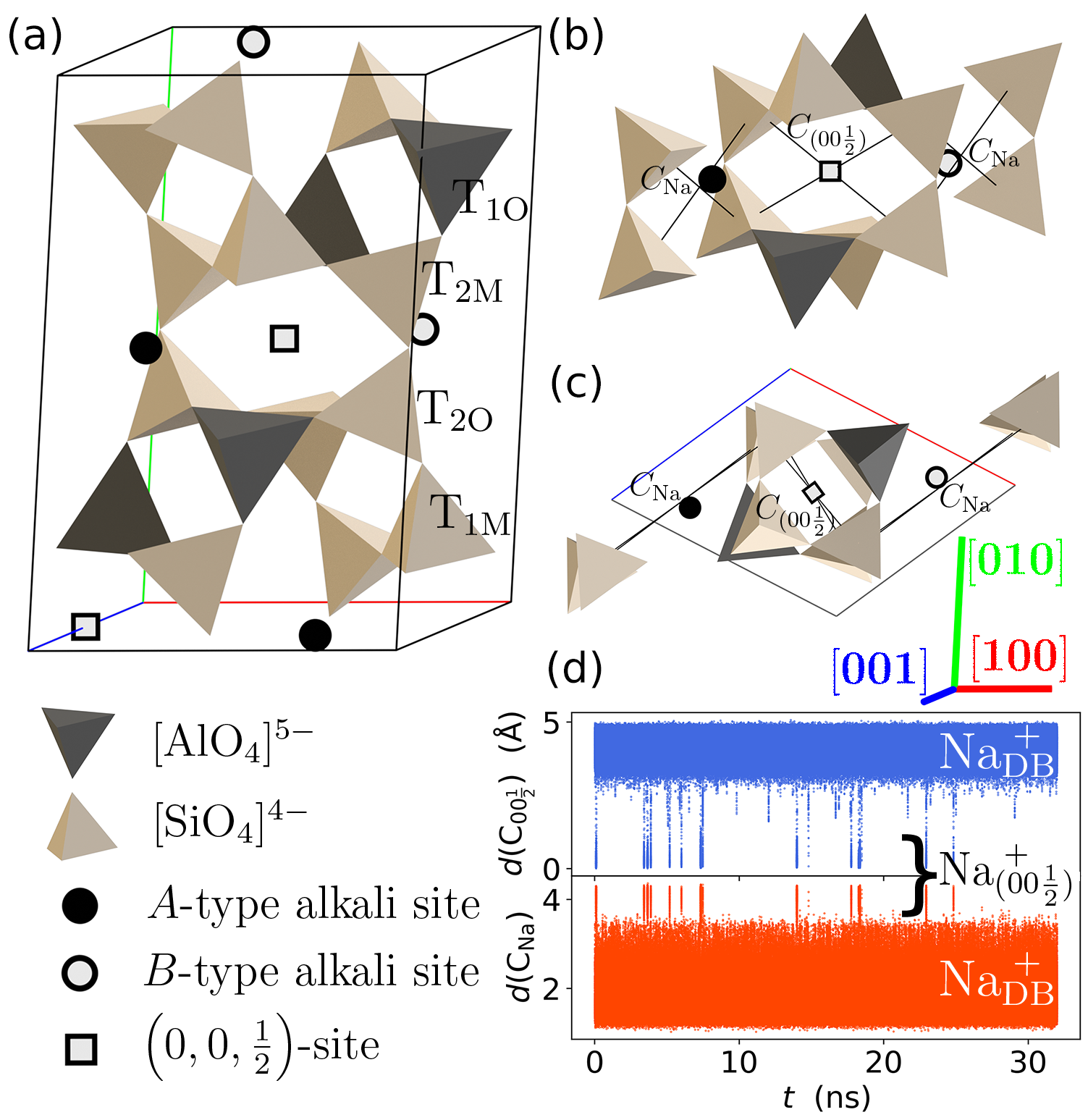
    }}
    \caption{In (\(\mathbf{a}\)) the conventional unit cell of Albite is shown with the \(A\)- and \(B\)-type alkali lattice sites and the \((0,0,\tfrac{1}{2})\)-sites. (\textbf{b}-\textbf{c}) show how these sites can be approximately located by the geometric centers of their 4 surrounding \(\textrm{T}_2\)-tetrahedra, which coincide with the intersection of the quadrilateral diagonals in the relaxed cell (and only there). In (\textbf{d}) the distances of an interstitial to those geometric centers is shown during an MD simulation at \SI{1000}{\kelvin}. They allow us to discern between \ch{Na^+_{DB}} and \ch{Na^+_{\((00\tfrac{1}{2})\)}}-type defects. %
    }   \label{fig:Cell_and_detection}
\end{figure}

\begin{figure}[h]
    \centering
    \makebox[\columnwidth][c]{\includegraphics[width=1.0\columnwidth]{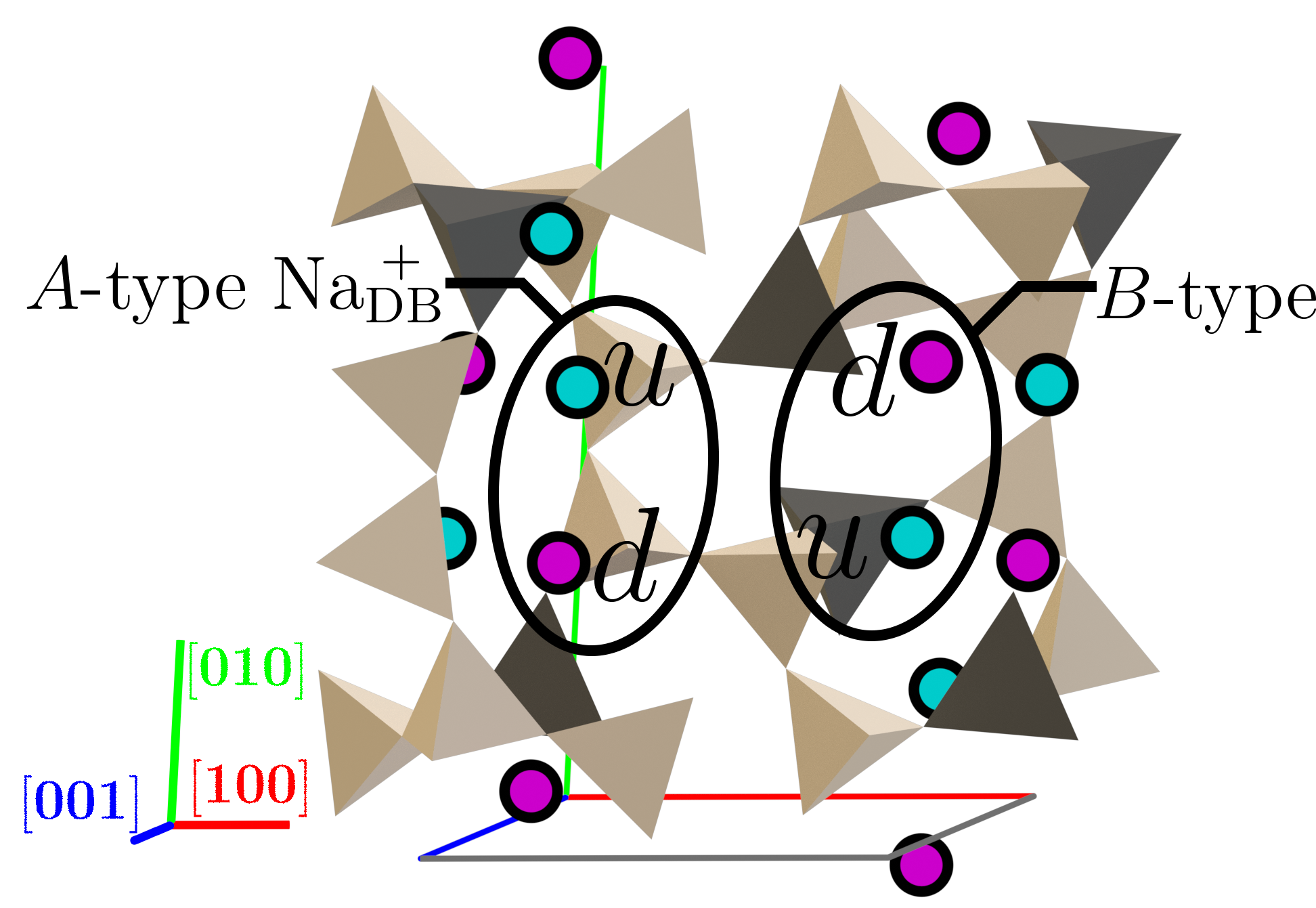
    }}
    \caption{Possible \(u\)- and \(d\)-states of a tracer ion in the ground state of the dumbbell defect \ch{Na^+_{DB}} in cyan and magenta. We differentiate between \(A\)- and \(B\)-type alkali sites that are shown centered in Fig.~\ref{fig:Defects_and_paths}~(a). 
    }
    \label{fig:updownfunk}
\end{figure}

The conventional unit cell of Albite (shown in Fig.~\ref{fig:Cell_and_detection}~a) consists of 16 corner-sharing aluminosilicate tetrahedra in a C\(\overline{1}\) space group. After considering symmetry operations - of which there are one translation and one inversion - only 4 unrelated tetrahedra remain. They are called \(\mathrm{T}_{1\mathrm{O}}\), \(\mathrm{T}_{1\mathrm{M}}\), \(\mathrm{T}_{2\mathrm{O}}\) and \(\mathrm{T}_{2\mathrm{M}}\). The defining feature of ordered Na-feldspar (Albite) is that aluminum occupies the \(\mathrm{T}_{1\mathrm{O}}\)-sites.

While the four alkali sites are also related by the same symmetry operations, for diffusion it is necessary to differentiate between non-inverted (\(A\)-type) and inverted (\(B\)-type) alkali-sites. A center of inversion coincides with the \((0,0,\tfrac{1}{2})\)-site illustrated in Fig.~\ref{fig:Cell_and_detection}. Interstitials are observed to occur at alkali sites in the form of dumbbells (see Fig.~\ref{fig:updownfunk}) and at \((0,0,\tfrac{1}{2})\)-sites as conventional interstitials.

To determine the positions \(\mathbf{r}_\mathrm{I}\) and \(\mathbf{r}_\mathrm{V}\) of the point defects - and with the goal of also having a lattice description of diffusion - we devised the following procedure. 

We first note that the location of an  alkali site can be very well approximated by the geometric center \(C_\mathrm{Na}\) of the four surrounding aluminosilicate tetrahedra, which are of type \(\mathrm{T}_{2\mathrm{O}}\) and \(\mathrm{T}_{2\mathrm{M}}\). This observation is illustrated in Fig.~\ref{fig:Cell_and_detection}~(b-c). The distance of the actual position of a sodium ion to its corresponding \(C_\mathrm{Na}\) is \SI{0.4}{\angstrom} in the relaxed system and by identifying sodium atoms to lattice sites with the closest \(C_\mathrm{Na}\), it is possible to uniquely match them even at the highest temperatures.

A dumbbell defect \ch{Na^+_{DB}} (depicted in Fig.~\ref{fig:updownfunk}) can then be located at a lattice site, which has two sodium ions closest to its \(C_\mathrm{Na}\). Moreover, of those two ions we identify the ion that is further away from its \(C_\mathrm{Na}\) as 'the interstitial' \ch{Na^+_{I}}. During a simulation the two occupants of the dumbbell vibrate around the alkali lattice site and we observe that the identity of \ch{Na^+_{I}} frequently switches between the two, however, during a diffusion event \ch{Na^+_{I}} is appropriately identified as the travelling ion. The procedure is independent of reference frame, and it is sufficient to accompany \ch{Na^+_{I}} as it travels through the crystal and record the interstitial position \(\mathbf{r}_\mathrm{I}\).

Using the distance \(d(C_{\ch{Na}})\) of \ch{Na^+_{I}} to \(C_\mathrm{Na}\) allows in principle to discern whether an \ch{Na^+_{I}} is in an \ch{Na^+_{\((00\tfrac{1}{2})\)}}-state already, as \(d(C_{\ch{Na}})\) will be invariably higher than possible in an \ch{Na^+_{DB}}-state. However, we cannot say at which \((0,0,\tfrac{1}{2})\)-site exactly the interstitial is located as there exist two closest \((0,0,\tfrac{1}{2})\)-sites at a similar distance from any alkali lattice site (see e.g.~Fig.~\ref{fig:Defects_and_paths}~d).

Therefore we also explicitly locate \ch{Na^+_{\((00\tfrac{1}{2})\)}} using a similar scheme. An occupied \((0,0,\tfrac{1}{2})\)-site can be approximated by the geometric center \(C_{(00\frac{1}{2})}\) of its four surrounding \(\mathrm{T}_{2\mathrm{O}}\) and \(\mathrm{T}_{2\mathrm{M}}\) tetrahedra shown in Fig.~\ref{fig:Cell_and_detection}~(b-c). We then assign the \ch{Na^+_{\((00\tfrac{1}{2})\)}} to the site with the closest \(C_{(00\frac{1}{2})}\). In Fig.~\ref{fig:Cell_and_detection}~(d) we plotted the distances of \ch{Na^+_{I}} to its alkali-lattice and \((0,0,\tfrac{1}{2})\)-sites during an MD simulation at \SI{1000}{\kelvin}. Intervals of high \(d(C_\mathrm{Na})\) or low \(d(C_{00\frac{1}{2}})\) both correctly identify an \ch{Na^+_{\((00\tfrac{1}{2})\)}} defect. The latter measure, however, fluctuates less, when an \((0,0,\tfrac{1}{2})\)-site is occupied and is therefore preferred. %

In its ground state, the dumbbell defect is oriented along the \([010]\) axis and therefore consists of a sodium that is closer to the alkali site in the positive \([010]\) direction. This is illustrated in Fig.~\ref{fig:updownfunk}. Also shown are that there are two possible states of the \ch{Na^+} inside the dumbbell. We denote them as \(u\)-type \ch{Na^+} and \(d\)-type \ch{Na^+}. %
Due to the inversion symmetry between \(A\)-type and \(B\)-type alkali sites, \(u\) and \(d\) states are also inverted relative to our fixed reference frame. %
To interpret the migration of a tracer atom, in particular its correlation, it is important to keep track of the \(u\) and \(d\) states of the \ch{Na^+} in the dumbbell. This was done by simply comparing the relative coordinates in the \([010]\) direction of the \ch{Na^+} in the dumbbell and labelling them accordingly. %

Finally, a vacancy \ch{V^-_{Na}} is identified by the absence of a sodium ion close to the \(C_\mathrm{Na}\) of an alkali-site.

\subsection{Lattice Description of Diffusion}
Diffusion in solids can be viewed as resulting from a sequence of jumps between different sites with jump rates
\begin{equation}
    \Gamma = \Gamma_0 \exp{\left(\frac{-\Delta G_a}{k_{\rm B}T}\right)}.
\end{equation}
Here, \(\Gamma_0\) is a prefactor and \(\Delta G_a\) is the activation energy of the defect migration between sites. Both the prefactor and the activation energy depend on the specific type of jump. Since the interstitial can occur in two states, we have to consider four reactions: \(\ch{Na^+_{DB}}\rightarrow\ch{Na^+_{DB}}\), \(\ch{Na^+_{DB}}\rightarrow\ch{Na^+_{\((00\tfrac{1}{2})\)}}\), \(\ch{Na^+_{\((00\tfrac{1}{2})\)}}\rightarrow\ch{Na^+_{DB}}\) and \(\ch{Na^+_{\((00\tfrac{1}{2})\)}}\rightarrow\ch{Na^+_{\((00\tfrac{1}{2})\)}}\). Furthermore, for each reaction there exist different pathways that - as we will find - have very differing rates.
To describe them we need to start with the lattice. Next, we will first give a qualitative overview of the difference jump processes and later discuss them quantitatively. 

\begin{figure}[h]
    \centering
    \makebox[\columnwidth][c]{\includegraphics[width=1\columnwidth]{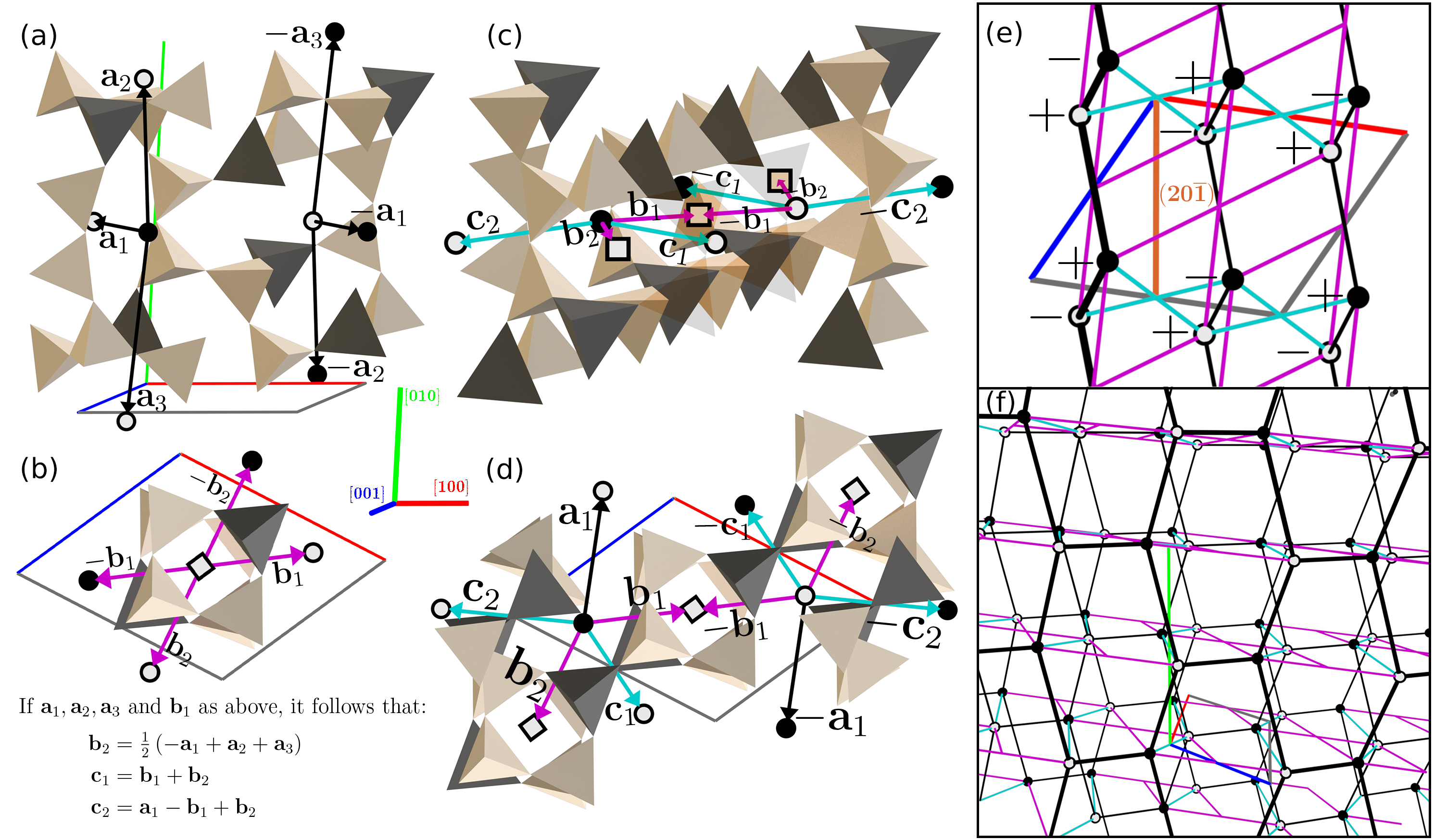
    }}
    \caption{Diffusion in the alkali-feldspar lattice. (\(\mathbf{a}\)-\(\mathbf{d}\)) illustrate the diffusion pathways that connect \(A\)-type alkali sites (black spheres), \(B\)-type alkali sites (open spheres) and \((0,0,\tfrac{1}{2})\)-sites (open squares). Only selected aluminosilicate tetrahedra are drawn, which show how \(\mathbf{a}_1\) passes through a 10-membered ring, \(\mathbf{a}_2\) and \(\mathbf{a}_3\) through a 6-membered ring, \(\mathbf{b}_1\), \(\mathbf{b}_2\) and \(\mathbf{c}_1\) through one 8-membered ring and \(\mathbf{c}_2\) through another 8-membered ring. (\(\mathbf{e}\)-\(\mathbf{f}\)) show two views of the lattice that is generated by the seven vectors \(\mathbf{a}_1\), \(\mathbf{a}_2\), \(\mathbf{a}_3\), \(\mathbf{b}_1\), \(\mathbf{b}_2\) \(\mathbf{c}_1\), and \(\mathbf{c}_2\), with the front-most \((20\overline{1})\)-honeycomb emphasized. The \((0,0,\tfrac{1}{2})\)-sites are not drawn, they coincide with magenta crossings (b).
    }   \label{fig:Defects_and_paths}
\end{figure}

The most important type of interstitial sodium migration occurs between alkali lattice sites through the reaction \(\ch{Na^+_{DB}}\rightarrow\ch{Na^+_{DB}}\). The primary paths for this reaction connect an alkali site to three neighbours, as illustrated in Fig.~\ref{fig:Defects_and_paths}~(a). On the most frequented path, denoted by \(\mathbf{a}_1\), the interstitial jumps through a 10-membered ring. This path keeps the defect in the \((010)\)-plane. To travel along \(\perp\!\!(010)\), the defect has to take a path along \(\mathbf{a}_2\) or \(\mathbf{a}_3\), which takes the defect through a six-membered ring. 

Since \( \left\{ \pm \mathbf{a}_1, \pm \mathbf{a}_2, \pm \mathbf{a}_3\right\}\) connect a site of type \(A/B\) to three sites of type \(B/A\), a honeycomb-like pattern emerges. In Fig.~\ref{fig:Defects_and_paths}~(f) this lattice is illustrated in a perspective view. A top view of the same lattice is shown in Fig.~\ref{fig:Defects_and_paths}~(e), in which it is apparent that the honeycomb generated by \( \left\{ \pm \mathbf{a}_1, \pm \mathbf{a}_2, \pm \mathbf{a}_3\right\}\) does not connect all alkali sites and that in fact, by restricting an interstitial to these paths it would be constrained to the \((20\overline{1})\)-plane.

Interstitials do, however, travel also between the \((20\overline{1})\)-honeycombs. To describe such paths we first introduce lattice directions that describe reactions in which the interstitial enters \((0,0,\tfrac{1}{2})\)-sites, \(\ch{Na^+_{DB}}\rightarrow\ch{Na^+_{\((00\tfrac{1}{2})\)}}\), and exits them, \(\ch{Na^+_{\((00\tfrac{1}{2})\)}}\rightarrow\ch{Na^+_{DB}}\). The corresponding pathways go along the vectors \(\mathbf{b}_1\) and \(\mathbf{b}_2\), as illustrated in Fig.~\ref{fig:Defects_and_paths}~(b). 

\newpage

\onecolumngrid

\twocolumngrid

\noindent Starting from an \ch{Na^+_{DB}}-state there are two neighbouring \((0,0,\tfrac{1}{2})\)-sites and from an \((0,0,\tfrac{1}{2})\)-site, an alkali atom can travel along \(\pm\mathbf{b}_1\) and \(\pm\mathbf{b}_2\) to four distinct alkali lattice sites. Since the vector \(\mathbf{b}_2 = \frac{1}{2} \left[-\mathbf{a}_1 + \mathbf{a}_2 + \mathbf{a}_3\right]\) is contained in the \((20\overline{1})\)-honeycomb, the path \(\ch{Na^+_{DB}}\rightarrow\ch{Na^+_{\((00\tfrac{1}{2})\)}}\rightarrow\ch{Na^+_{DB}}\) along twice \(\mathbf{b}_2\) does not exit a \((20\overline{1})\)-honeycomb. However, the five other such paths that contain \(\mathbf{b}_1\) do.

While all diffusion processes can now be described to happen along a path of some combination of the vectors introduced above, two of them are sufficiently relevant to be denominated in addition. They describe the two other \(\ch{Na^+_{DB}}\rightarrow\ch{Na^+_{DB}}\) events in the \((010)\)-plane. As illustrated in Fig.~\ref{fig:Defects_and_paths}~(c-d), they occur along \(\mathbf{c}_1 = \mathbf{b}_1 + \mathbf{b}_2\) and \(\mathbf{c}_2 = \mathbf{a}_1 - \mathbf{b}_1 + \mathbf{b}_2\). The notation is supposed to distinguish between the combined event \(\ch{Na^+_{DB}}\rightarrow\ch{Na^+_{\((00\tfrac{1}{2})\)}}\rightarrow\ch{Na^+_{DB}}\) along (\(\mathbf{b}_1\), then \(\mathbf{b}_2\)) and the direct path along \(\mathbf{c}_1\). While the result of both reactions are identical, the former has a much smaller rate compared to the latter.

Since vacancies can only travel between alkali lattice sites by \(\ch{V^-_{Na}}\rightarrow\ch{V^-_{Na}}\), they do not have an associated rate along \(\mathbf{b}_1\) or \(\mathbf{b}_2\). We will see later that they prefer traveling along \(\mathbf{c}_1\) or \(\mathbf{a}_1\) and almost never move in \(\perp\!\!(010)\).

\section{Results and Discussion}\label{sec:results}
\subsection{Defect Diffusion}
\subsubsection{Jump Rates of Point Defects}

\begin{figure}[t]
    \centering
    \makebox[\columnwidth][c]{\includegraphics[width=1\columnwidth]{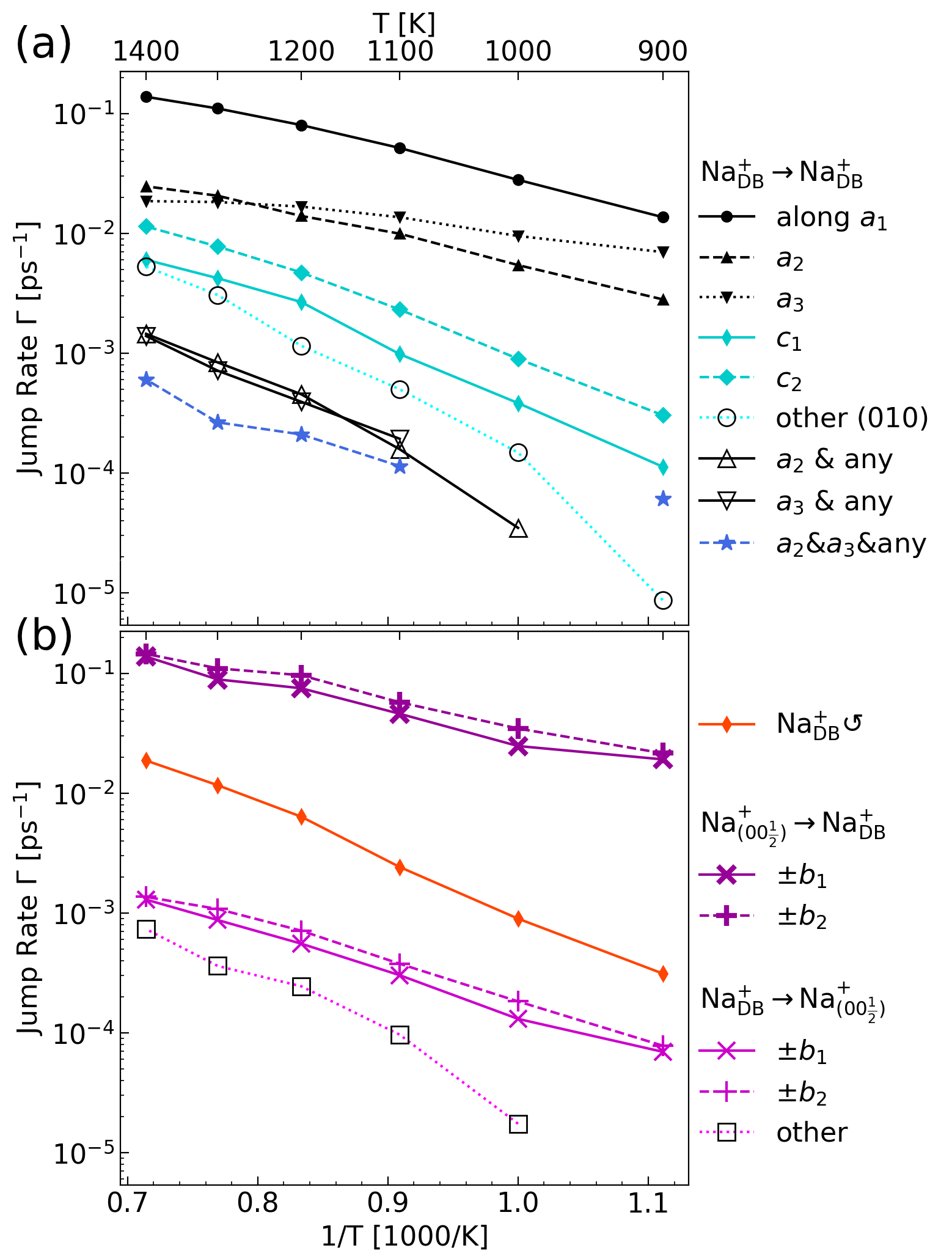
    }}
    \caption{Jump rates along lattice directions (refer to Fig.~\ref{fig:Defects_and_paths}~[a-d]) measured for both \(A\)-type and \(B\)-type sites collapsed into a single rate. In \(\mathbf{(a)}\) only reactions between alkali sites are considered. Higher order terms are summarized separately for events in \((010)\), \(\mathbf{a}_2\) or \(\mathbf{a}_3\) and any additional jump in \((010)\) as well as double \([010]\)-jumps and any or no additional jump in \((010)\) (blue stars). In \(\mathbf{(b)}\) exit rates from an \((0,0,\tfrac{1}{2})\)-site to an alkali site along either positive or negative \(\mathbf{b}_2\) and \(\mathbf{b}_1\) are shown together. An exit along e.g.~\(+\mathbf{b}_2\) specifically would have half the rate shown here. %
    The reaction \(\ch{Na^+_{\((00\tfrac{1}{2})\)}}\rightarrow\ch{Na^+_{\((00\tfrac{1}{2})\)}}\) was not observed, i.e.~we measured a rate of zero.
    }   \label{fig:Hops_Int}
\end{figure}

In Fig.~\ref{fig:Hops_Int} the jump rates of the interstitial along the directions defined in Fig.~\ref{fig:Defects_and_paths} are plotted over temperature. We observe that jumps along \(\mathbf{a}_1\), \(\mathbf{a}_2\) and \(\mathbf{a}_3\) are always the most frequent. The \(\mathbf{a}_1\) jump may be described as the dominant reaction. For the purpose of diffusion the significance of \(\mathbf{a}_1\) jumps by themselves is, however, limited,  because they only allow the defect to switch from some \(A\) site to \(B\) and back from \(B\) to the same \(A\) site. Any substantial motion of the defect requires additional jumps.

The \(\mathbf{a}_1\), \(\mathbf{a}_2\) and \(\mathbf{a}_3\) jumps keep the defect on the \((20\overline{1})\)-honeycombs. The major modes of transport between honeycombs are \(\mathbf{c}_2\) and \(\mathbf{c}_1\). Other possible channels within the \((010)\)-plane become relevant only at the highest temperatures. We interpret these as higher order events in which the interstitial either kicks out a lattice atom which jumps further or as an event in which the interstital jumps by two sites at once. Such higher order jumps can also contain a move in the \([010]\) direction  
(\(\mathbf{a}_2\) or \(\mathbf{a}_3\)) and on rare occasion even both \(\pm\mathbf{a}_2\) and \(\mp\mathbf{a}_3\) in one event (two \([010]\)-jumps). Overall, however, higher order events in the \([010]\) direction have a very minor rate compared to those in the \((010)\) plane, which in turn also become a relevant contribution to sodium transport only above \SI{1200}{\kelvin}.

The rates of events \(\ch{Na^+_{DB}}\rightarrow\ch{Na^+_{\((00\tfrac{1}{2})\)}}\) are shown in Fig.~\ref{fig:Hops_Int}~(b). Compared to transitions between alkali sites, the interstitial enters an \((0,0,\tfrac{1}{2})\)-site only infrequently. The reaction to exit the site, \(\ch{Na^+_{\((00\tfrac{1}{2})\)}}\rightarrow\ch{Na^+_{DB}}\), on the other hand has a rate similar to that of the \(\mathbf{a}_1\)-transition for both paths along \(\pm\mathbf{b}_1\) or \(\pm\mathbf{b}_2\). We can therefore expect the residence time of an interstitial at an \((0,0,\tfrac{1}{2})\)-site to be similar to its residence time in a dumbbell state. %

\begin{figure}[t]
    \centering
    \makebox[\columnwidth][c]{\includegraphics[width=1\columnwidth]{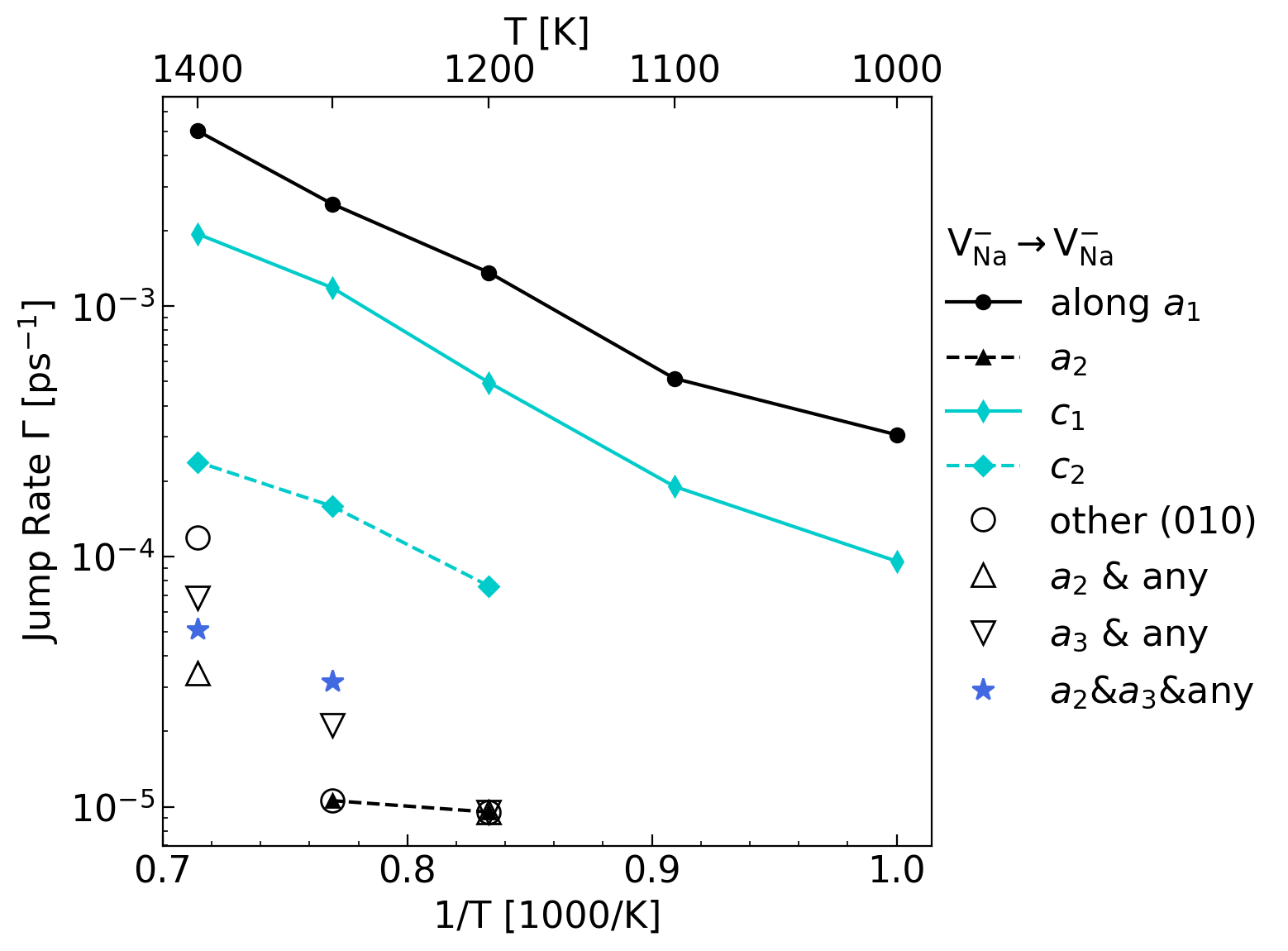
    }}
    \caption{Jump rates of the vacancy along jump vectors (see Fig.~\ref{fig:Defects_and_paths}). The vacancy only moves from an alkali site to another alkali site as opposed to the interstitial (cf.~Fig.~\ref{fig:Hops_Int}).
    }   \label{fig:Hops_Vac}
\end{figure}

The large differences between the jump rates in the different directions can be qualitatively explained by considering that the dumbbell is preferentially oriented in the [010] direction. %
From this state it seems plausible that the upper or lower \ch{Na^+} can directly transition through the \([010]\)-oriented rings in the \(\mathbf{a}_1\) or \(\mathbf{c}_2\) directions and the paths in the \([010]\) direction through either \(\mathbf{a}_2\) or \(\mathbf{a}_3\) are easily accessible from either the upper or lower \(\ch{Na^+}\) in the dumbbell. To move through the ring in \(\mathbf{b}_1\), \(\mathbf{b}_2\) or \(\mathbf{c}_1\) on the other hand, the dumbbell must first transition to a configuration in which its axis lies more parallel to the \((010)\)-plane, before it can proceed with a jump through the eight-membered ring since the major axis of the latter also lies in the \((010)\)-plane. In short, the activation volume of migrations along \(\mathbf{a}_1\), \(\mathbf{c}_2\), \(\mathbf{a}_2\) and \(\mathbf{a}_3\) is likely smaller than for \(\mathbf{b}_1\), \(\mathbf{b}_2\) or \(\mathbf{c}_1\). This may offer also an explanation for the differing jump rates of interstitial and vacancy, as much less preference for \([010]\)-oriented rings should be expected for the vacancy, and none are indeed visible in Fig.~\ref{fig:Hops_Vac}.

Since feldspar becomes more monoclininc with increasing temperatures, which was observed also for the NNP that is used here~\cite{gorfer2024structure}, we may interpret the observation that the rates along \(\mathbf{a}_2\) and \(\mathbf{a}_3\) approach each other with increasing temperature as an effect of the \((010)\) pseudo-mirror-plane becoming closer to a true mirror-plane with increasing temperature. The closer the \((010)\)-plane becomes to a true mirror, the more similar \(\mathbf{a}_2\) and \(\mathbf{a}_3\) become.

The dumbbell defect is able to rotate in a reaction we label as \(\ch{Na^+_{DB}}\circlearrowleft\), which changes the \(u/d\) state of the two constituent \ch{Na^+} into \(d/u\). This reaction has no influence on the diffusion coefficent of the interstitial, but it is highly relevant for \ch{Na^+} self-diffusion as it decreases correlations. This is explored more deeply in Section~\ref{sec:correlation_effects}, but essentially the diffusion of \ch{Na^+} in the \([010]\) direction is slowed without dumbbell rotations since the \(u/d\) \ch{Na^+} can only jump once along  \(\mathbf{a}_2/\mathbf{a}_3\) and back again.

The reaction \(\ch{Na^+_{\((00\tfrac{1}{2})\)}}\rightarrow\ch{Na^+_{\((00\tfrac{1}{2})\)}}\) never occurred during the simulation, i.e.~we measured a rate of zero. From a geometric point of view this is somewhat surprising. It can be seen from Fig.~\ref{fig:Defects_and_paths}~(c) that neighbouring \((0,0,\tfrac{1}{2})\)-sites lie in an open channel along the \([001]\)-direction. It seems that travel along this path is unstable with respect to falling into one of the alkali sites.

In Fig.~\ref{fig:Hops_Vac} the jump rates for the vacancy are shown. The path along \(\mathbf{a}_1\) is again the most frequently travelled, although its rate is two orders of magnitude smaller than the corresponding rate for the interstitial defect. Jumps along \(\mathbf{c}_1\) come in second. Since these two reactions dominate, we should expect the vacancy to prefer travelling in the \([101]\) direction since \((\mathbf{a}_1 - \mathbf{c}_1)\) coincides with this direction. In the case of K-rich feldspar, an experiment of Petrisheva et al.~\cite{Petrishcheva2014} showed that Na-K interdiffusion is fastest along \([101]\). The path along \(\mathbf{c}_2\) becomes active at \SI{1200}{\kelvin} but remains of minor importance compared to \(\mathbf{a}_1\) or \(\mathbf{c}_1\). Jumps in the \([010]\) direction are even more rare.

\subsubsection{Diffusion Coefficients of Point Defects}
\label{Sec:DiffusionInterstitialDefects}

\begin{figure}[t]
    \centering
    \makebox[\columnwidth][c]{\includegraphics[width=1\columnwidth]{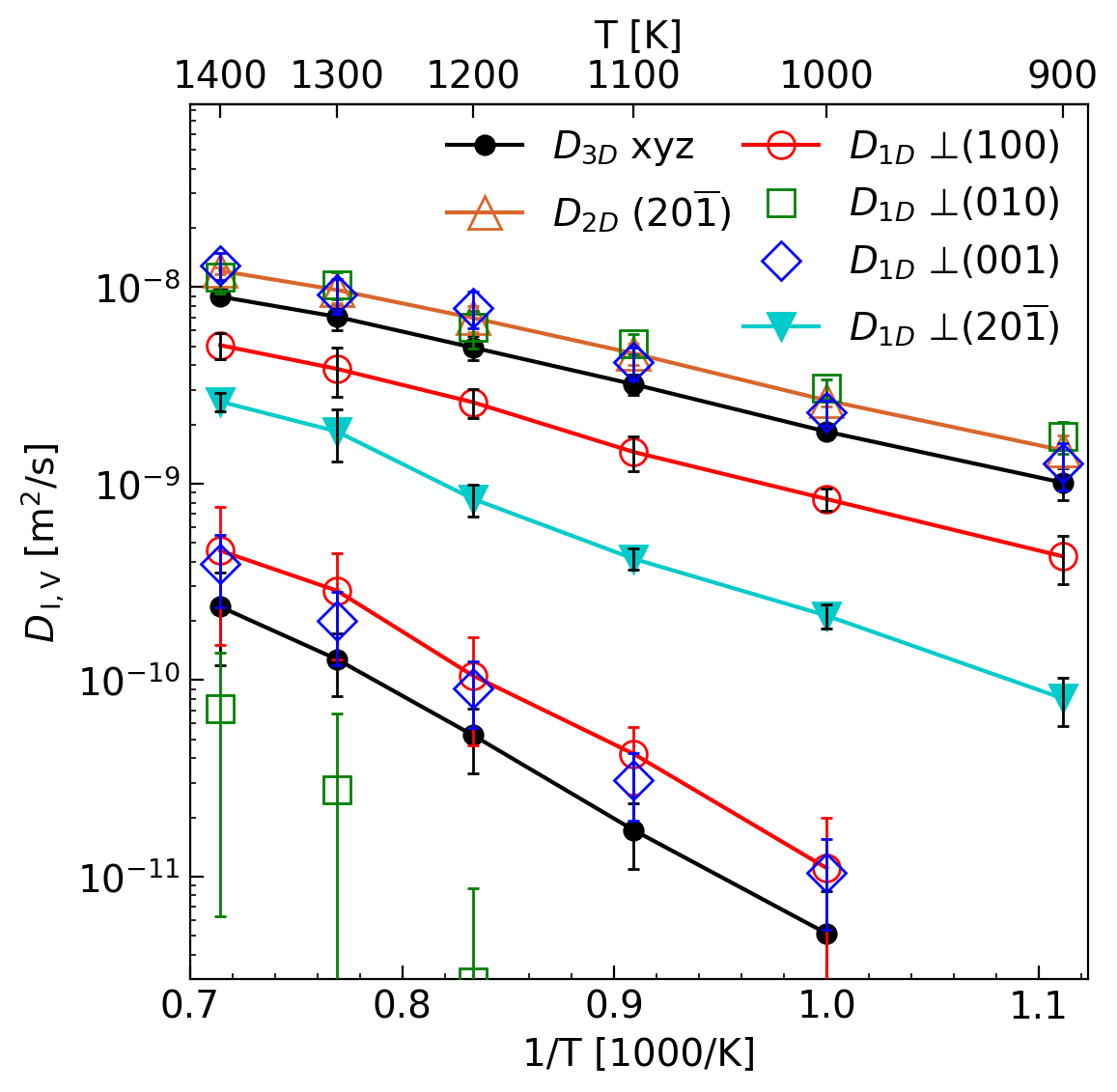
    }}
    \caption{Diffusion coefficients of the interstitial \(D_{\mathrm{I}}\) (upper 6) and of the vacancy \(D_{\mathrm{V}}\) (lower 4) measured over all dimensions \(D_{3D}\) and projected onto two \(D_{2D}\) or one dimension \(D_{1D}\).
    }   \label{fig:Dcoeff_combo}
\end{figure}

The diffusion coefficients of interstitial and vacancy defects computed from the respective mean square displacements (Eq.~\ref{Eq:D_I}) are shown in Fig.~\ref{fig:Dcoeff_combo}.

We observe that the preference of the interstitial to travel along \(\mathbf{a}_1, \mathbf{a}_2\) and \(\mathbf{a}_3\) results in a highly anisotropic diffusivity. The 2D-diffusion coefficient on the \((20\overline{1})\)-plane is about 12 times larger at \SI{1000}{\kelvin} than the 1D diffusion coefficient along \(\perp\!\!(20\overline{1})\). The difference reduces to 8 times at \SI{1400}{\kelvin}, a fact which we can connect to the higher slope of jump rates along \(\mathbf{c}_1\) and \(\mathbf{c}_2\) compared to \(\mathbf{a}_2\) and \(\mathbf{a}_3\) in Fig.~\ref{fig:Hops_Int}~(a).

The diffusion coefficients for the vacancy are significantly smaller than for the interstitial, which is in line with interpretations of experiments of Ca-rich \cite{Behrens1990} and K-rich feldspar \cite{wilangowski2015_proceeding}. If we compare the three dimensional coefficient \(D_{3D}\), the difference varies from around a factor of 27 at \SI{1000}{\kelvin} to a factor of 14 at \SI{1400}{\kelvin}. Even the fastest mode, the diffusion in the \([101]\)-direction which is similar in magnitude to the \(\perp(100)\) and \(\perp(001)\) directions, does not come close to the slowest mode of interstitial transport. %
Very striking is the sluggish diffusion in the \(\perp(010)\) direction.

\subsection{Self-Diffusion of Na-ions and Correlation Effects}

In the last section we considered the diffusion of point defects. Building upon this, let us next investigate the mobility of sodium ions. To relate the diffusion coefficient of the sodium ion to the diffusion coefficients of the defects, correlation effects must be taken into account.

\subsubsection{Correlation Effects on Jump Rates}\label{sec:correlation_effects}

As discussed above, the \(\ch{Na^+_{DB}}\rightarrow\ch{Na^+_{DB}}\) reaction constitutes the primary migration mode of the interstitial defect (cf. Fig.~\ref{fig:Hops_Int}). %
The dumbbell consists of a \(u\)-type Na that is closer to the neighbouring alkali site in the \(\mathbf{a}_2\) direction and a \(d\)-type Na that is closer to the alkali site in the \(\mathbf{a}_3\) direction (see Fig.~\ref{fig:updownfunk}).
The most important consequence of this configuration is that if the tracer atom sits at the \(u\)/\(d\) site, we may expect that it is much more likely to move in the \(\mathbf{a}_2\)/\(\mathbf{a}_3\) direction and not \(\mathbf{a}_3\)/\(\mathbf{a}_2\), as in the latter there are one or even two \ch{Na^+} in its way.

Let us quantify this notion using conditional probabilities. Given any \(\ch{Na^+_{DB}}\rightarrow\ch{Na^+_{DB}}\) reaction of the defect in a particular direction (\(\mathbf{a}_1\), \(\mathbf{a}_2\) ...), we determined the frequency of the four possible tracer moves \(\left\{(u\rightarrow u),(d\rightarrow d), (u\rightarrow d), (d\rightarrow u) \right\}\). In the latter two reactions, which {\it invert} \(u\) and \(d\), the atoms trace the same path and because of microscopic reversibility these have the same conditional probability. %

\begin{figure}[h!]
    \centering
    \makebox[\columnwidth][c]{\includegraphics[width=1\columnwidth]{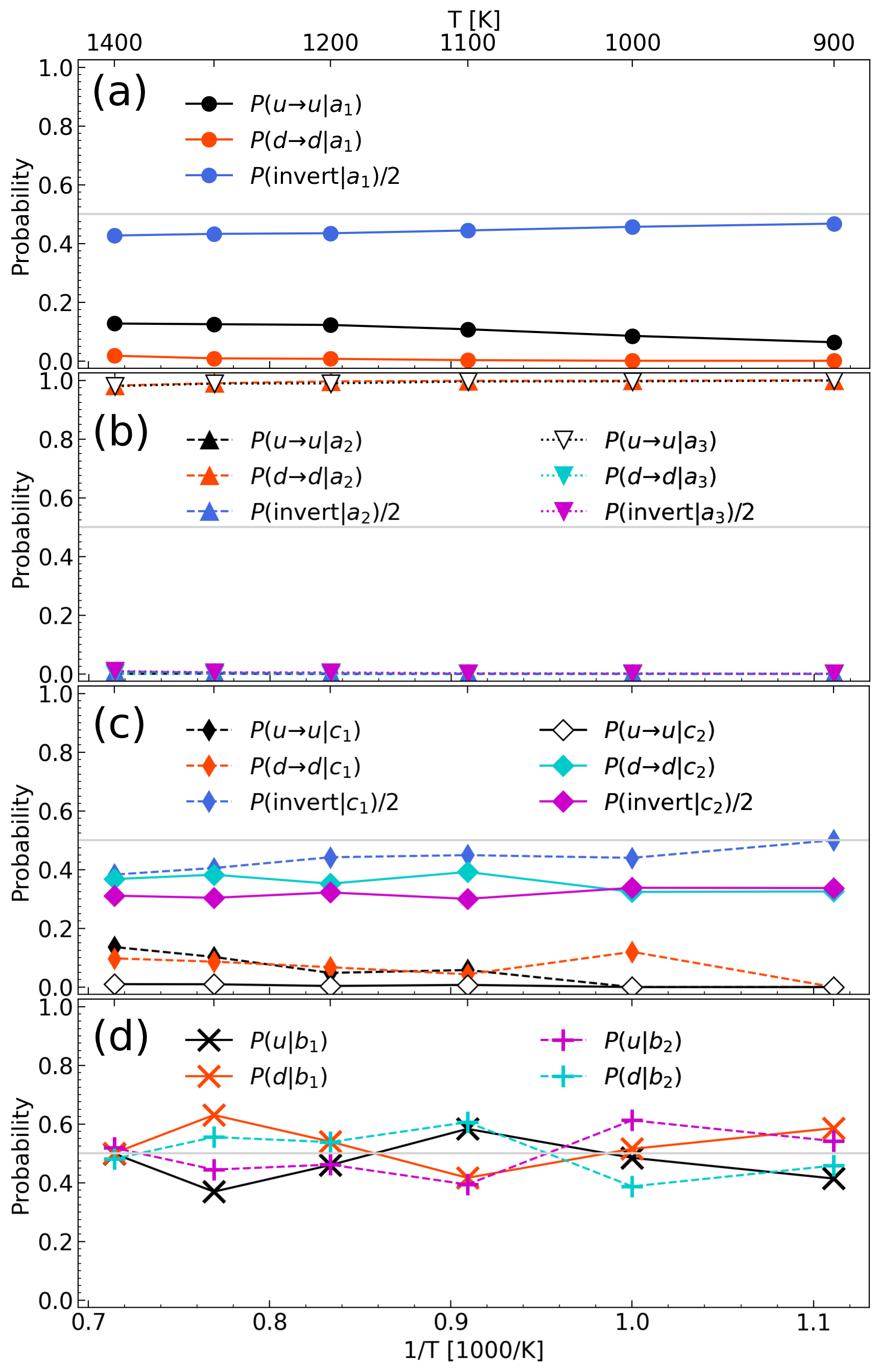}}
    \caption{ (\(\mathbf{a}\) - \(\mathbf{c}\)) Conditional probabilities of a tracer atom jumping from an \(u\) or \(d\) state to an \(u\) or \(d\) state if the interstitial defect performed an \(\ch{Na^+_{DB}}\rightarrow\ch{Na^+_{DB}}\) type jump. Jumps which invert \(u\) into \(d\) or \(d\) into \(u\) have the same probability. Their probability \(P(\mathrm{invert|\mathbf{a}_1})/2\) is plotted only once. \(\mathbf{(d)}\) Conditional probability of jumping into/out-of a \(u\) or \(d\) state if the defect performed a \(\ch{Na^+_{DB}}\rightarrow\ch{Na^+_{\((00\tfrac{1}{2})\)}}\) or \(\ch{Na^+_{\((00\tfrac{1}{2})\)}}\rightarrow\ch{Na^+_{DB}}\) type jump.
    }   \label{fig:Correlation_Probabilities}
\end{figure}

The resulting probabilities, shown in Fig.~\ref{fig:Correlation_Probabilities}~(b), confirm our intuition; if an \(\mathbf{a}_2\)/\(\mathbf{a}_3\) jump occurs, the motion of the tracer atom is almost entirely due to a jump from an \(u\)/\(d\) to the close \(u\)/\(d\) site. In the case of \(\mathbf{a}_1\) the difference in the probabilities is less extreme but a major amount of the jumps in this direction are of the inversion type \((u\rightarrow d)\) and \((d\rightarrow u)\). In this move, the tracer atom does not cross the \((010)\)-plane, so the possibility of the tracer to migrate in the \(\perp(010)\) direction in a future jump is as low as before the \(\mathbf{a}_1\). If we were to only consider the most frequent defect jumps \(\mathbf{a}_1\), \(\mathbf{a}_2\) and \(\mathbf{a}_3\), the mobility of the tracer atom in the \(\perp(010)\) direction would actually be almost entirely dependent on \(u\rightarrow u\) transitions in \(\mathbf{a}_1\). These only make up about 10\% of all \(\mathbf{a}_1\) transitions and therefore have a similar rate to the rotational transition of the dumbbell (cf.~Fig.~\ref{fig:Hops_Int}) which has a similar effect with regard to \(\perp(010)\) mobility.

Figure~\ref{fig:Correlation_Probabilities} implies that we can conceptualize four different jump rates per possible \(\ch{Na^+_{DB}}\rightarrow\ch{Na^+_{DB}}\) jump that sum up to give the defect jump rates of Fig.~\ref{fig:Hops_Int}. The jump rates are related by
\begin{equation*}
    \Gamma_{\mathbf{a}_1} P(u\rightarrow u|a_1) = \Gamma_{\mathbf{a}_1}^{u\rightarrow u},
\end{equation*}
where \(\Gamma_{\mathbf{a}_1}^{u\rightarrow u}\) is the rate of the \(\mathbf{a}_1\) by \(u\rightarrow u\) reaction and 
\begin{equation*}
    \Gamma_{\mathbf{a}_1}^{u\rightarrow u} + \Gamma_{\mathbf{a}_1}^{d\rightarrow d} + \Gamma_{\mathbf{a}_1}^{u\rightarrow d} + \Gamma_{\mathbf{a}_1}^{d\rightarrow u} =  \Gamma_{\mathbf{a}_1}.
\end{equation*}

\subsubsection{Self-Diffusion Coefficients and Correlation Factor}

\begin{figure}[h]
    \centering
    \makebox[\columnwidth][c]{\includegraphics[width=1\columnwidth]{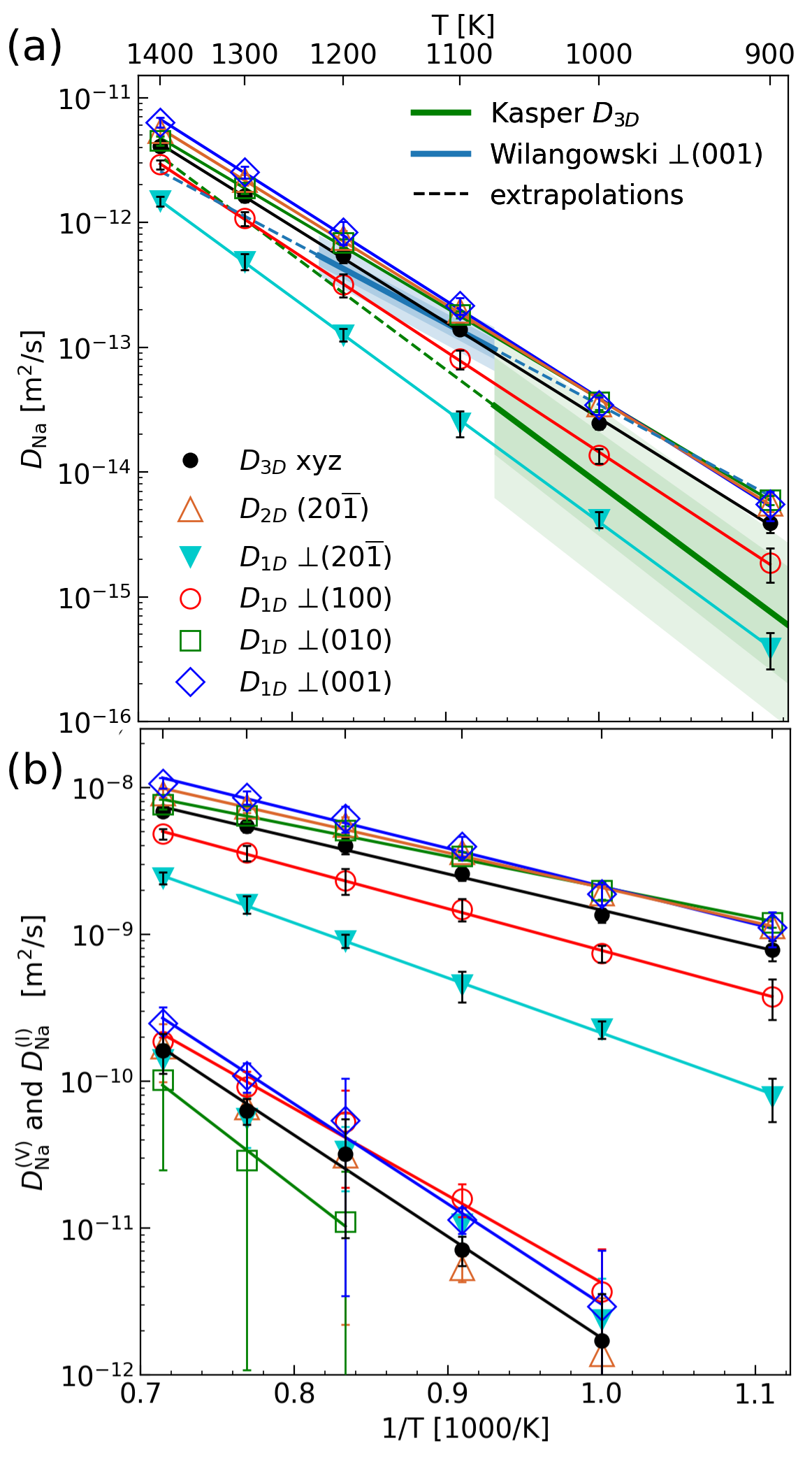
    }}
    \caption{ \((\mathbf{a})\) The self-diffusion coefficients determined over all dimensions \(D_{3D}\), in selected one dimensions \(D_{1D}\) and in the honeycomb plane \(D_{2D}\). Also shown are experimental data of Kasper \cite{Kasper1975} and Wilangowski and coworkers~\cite{wilangowski2015_proceeding}. In \((\mathbf{b})\) the sodium diffusivity per defect is shown. Upper 6 are \(D^{(\mathrm{I})}_{\ch{Na}}\) and lower 6 are \(D^{(\mathrm{V})}_{\ch{Na}}\). Straight lines are Arrhenius fits.
    }   \label{fig:Dcoeff_all}
\end{figure}

The self-diffusion coefficients of sodium ions computed according to Eq.~(\ref{eq:na_selfdiffusion}) are plotted in Fig.~\ref{fig:Dcoeff_all}~(a) together with experimental results of Wilangowski et al.~\cite{wilangowski2015_proceeding} and Kasper~\cite{Kasper1975}. Wilangowski used a K-rich (\(X_\text{K} = 0.85\)) sanidine and measured the Na tracer diffusion coefficient along \(\perp\!(001)\). Our result for the three dimensional coefficient matches Kasper's absorption experiments, at least within the mutual uncertainty. 
Our result for the one dimensional diffusion coefficient along \(\perp\!(001)\) aligns with the findings of Wilangowski's tracer studies from \SI{900}{\kelvin} to \SI{1100}{\kelvin} but deviates minimally above the common uncertainty range at and above \SI{1200}{\kelvin}. Since Wilangowski studied Na self-diffusion in K-rich feldspar, which is expected to hinder Na diffusion~\cite{Hergemoeller2017}, such a deviation is not unexpected. In Fig.~\ref{fig:Dcoeff_all}~(b) the sodium diffusivities per defect used to determine the sodium self diffusion coefficient are shown.

The considerations given for the diffusivity of interstitial defects in section~\ref{Sec:DiffusionInterstitialDefects} can essentially be transferred to \ch{Na^+}~self-diffusion, as the contribution of \(D^{(\mathrm{V})}_{\ch{Na}}\) to the total sodium diffusion coefficient is negligible. The high anisotropy is again visible in the two slow coefficients in \(\perp\!\!(100)\) and \(\perp\!\!(20\overline{1})\). This anisotropy matches with the Na-K interdiffusion experiments of Petrishcheva~\cite{Petrishcheva2014} where \(\perp\!\!(10\overline{1})\) was predicted to be slowest. 

Using \(D_{\mathrm{I}}\) and \(D^{(\mathrm{I})}_{\ch{Na}}\) (Fig.~\ref{fig:Dcoeff_combo} and Fig.~\ref{fig:Dcoeff_all}~b) we determined the correlation factors across temperature using Eq.~(\ref{Eq:Def_Correlationfactors}). Since the resulting correlation factors are independent of temperature, we present the \(T\)-average in Table~\ref{Tab:correlation_factors}.
The most striking result is the markedly low correlation factor in the \(\perp(010)\) direction. This reaffirms the notion that a sodium tracer atom is prevented from jumping in this direction because of its partner dumbbell ion, pictured in Fig.~\ref{fig:updownfunk} and is in accordance with the probabilities of Fig.~\ref{fig:Correlation_Probabilities}. %

\begin{table}[h]
    \centering
    \caption{Correlation factors for the interstitial defect.}
    \begin{tabular}{c|c c c c c c }
                        & 3D & \((20\overline{1})\) & \(\perp\!(20\overline{1})\) & \(\perp\!(100)\) & \(\perp\!(010)\) & \(\perp\!(001)\) \\\toprule
      \(f_\mathrm{I}\)   & 0.77 & 0.76 &   1.0 &   0.93 &   0.67 &   0.85  \\
      \(\Delta f_\mathrm{I}\)& 0.03 & 0.03 &   0.1 &   0.05 &   0.06 &   0.07  \\ \hline

    \end{tabular}
    \label{Tab:correlation_factors}
\end{table}

In the 1990 experiment of Behrens \cite{Behrens1990} the sodium diffusivity in plagioclase (\ch{An_{62}}) at \SI{1073.15}{\kelvin} and \SI{1173.15}{\kelvin} in the \(\perp\!\!(010)\) and \(\perp\!\!(001)\) direction was found to be identical within experimental uncertainty. Our in-silico study replicates this finding between 900 to \SI{1200}{\kelvin} but suggests that the \(\perp\!\!(001)\) direction is slightly faster than \(\perp\!\!(010)\). This result in turn does also align with the 1974 experiment of Giletti et al.~\cite{giletti_1974}, the 1983 measurements of direction dependent interdiffusion coefficients of Christoffersen and coworkers~\cite{christoffersen1983} and the 2014 study of Schaeffer et al.~\cite{Schaeffer2014Sodium} as well as with measurements of ionic conductivity \cite{ElMaanaoui2016} in high-K feldspar, which all show an anisotropy \(\perp\!\!(001) > \, \perp\!\!(010)\), although significantly more extreme. Considering that our data show that diffusion of \ch{Na^+} in Na-feldspar is already strongly influenced by correlation effects in the \(\perp\!\!(010)\) direction, if we assume that the dumbbell mechanism also dominates cation diffusion in these \ch{K^+}- or \ch{Ca^{2+}}-rich feldspars, it might be that these more extreme anisotropies originate in the considerably larger cations which we would expect to have an even higher influence on correlation. %

\subsubsection{Haven Ratios in Na-feldspar}

The Haven Ratio \(H_\mathrm{R}\) is a measurable quantity that relates the diffusion of the tracer atoms \(D_{\ch{Na}}\) to the charge diffusivity \(D_{\sigma}\). It fulfills
\begin{equation*}
    H_\mathrm{R} = \frac{D_{{\ch{Na}}}}{D_{\sigma}} = f \frac{a_{\ch{Na}}}{a_I},
\end{equation*}
where \(f\) is the correlation factor, \(a_{\ch{Na}}\) is the jump length of the \ch{Na^+} ions and \(a_I\) is the jump length of the charge.

For the jump mechanisms that we have observed, the charge moves the same distance as the tracer in almost all cases. Only in rare cases, the defect moves similarly to an interstitialcy mechanism \cite{mehrer_diffusion_2007} in which two ions travel simultaneously  (cf.~Fig.~\ref{fig:Hops_Int} (\(a_2\)~\&~any)). Hence, we can set \(\tfrac{a_{\ch{Na}}}{a_I} = 1\). This leaves the correlation factors of Tab.~\ref{Tab:correlation_factors} as our predictions of Haven Ratios in Na-feldspar.  Previously, Haven rations in alkali-feldspar have been approximated using the collinear interstitialcy mechanism in simple cubic crystals, which gives an \(H_\mathrm{R}\) equal to \(0.43\)~\cite{ElMaanaoui2016}.

\subsection{Characterization of the Interstitial}

\begin{figure}[h]
    \centering
    \makebox[\columnwidth][c]{\includegraphics[width=1\columnwidth]{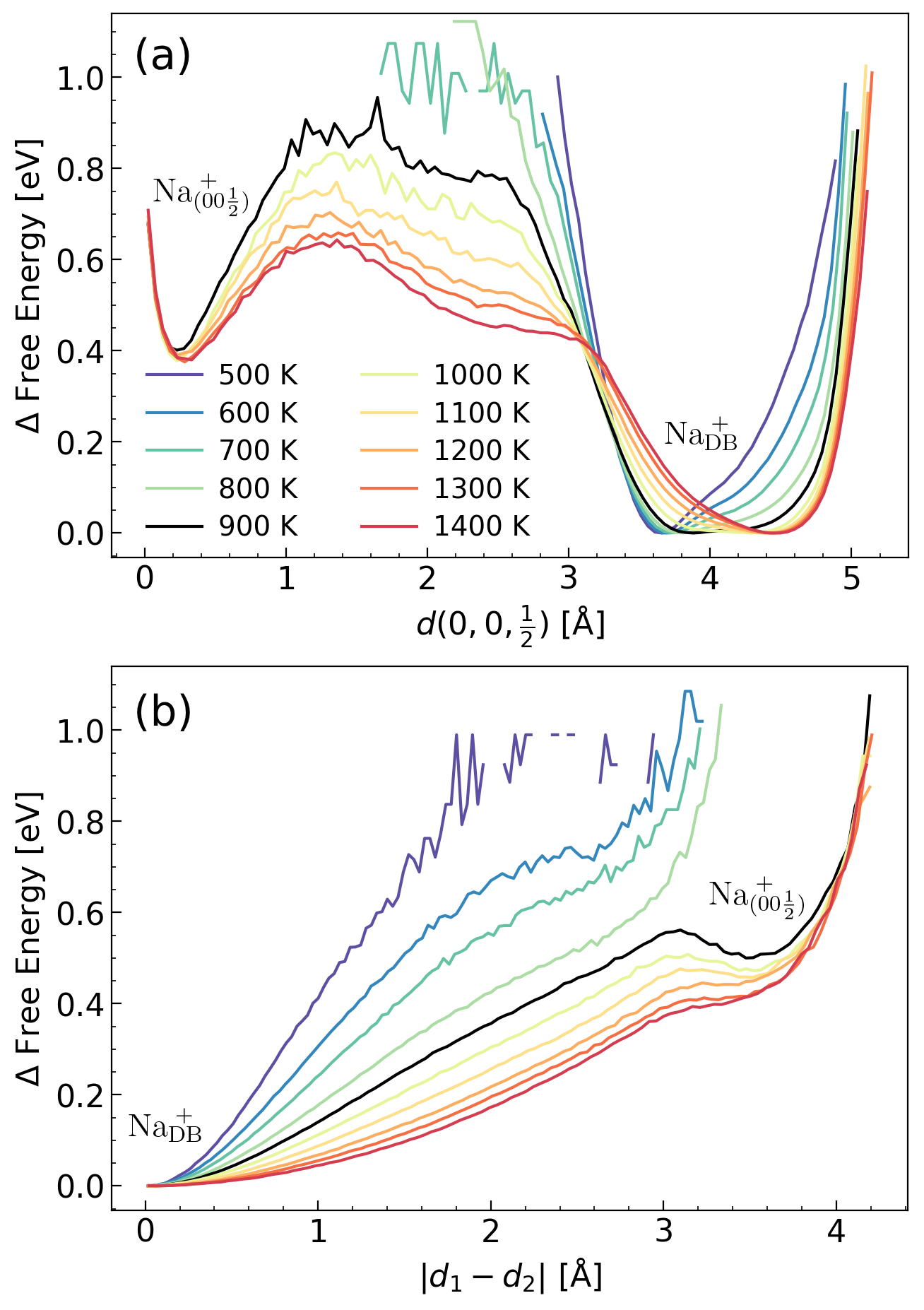
    }}
    \caption{Free energy profiles for the interstitial defect. In (\textbf{a}) the collective coordinate used to characterize the defect configuration is the distance to the closest \((0,0,\tfrac{1}{2})\)-site and in (\textbf{b}) it is the difference of the distances of the two sodium atoms in the dumbbell to their shared alkali lattice site. In both (\textbf{a}-\textbf{b}) the free energy has a double-well form, where the  lower well corresponds to a \ch{Na^+_{DB}} defect-state and the higher well to a \ch{Na^+_{\((00\tfrac{1}{2})\)}}. The reaction \(\ch{Na^+_{DB}}\rightarrow\ch{Na^+_{\((00\tfrac{1}{2})\)}}\) was not observed below \SI{900}{\kelvin}.
    }   \label{fig:site_dists}
\end{figure}

The difference in the rates of \(\ch{Na^+_{DB}}\rightarrow\ch{Na^+_{\((00\tfrac{1}{2})\)}}\) and its reverse \(\ch{Na^+_{\((00\tfrac{1}{2})\)}}\rightarrow\ch{Na^+_{DB}}\) indicate that the  \ch{Na^+_{DB}} defect configuration is the more stable one. This is confirmed by the free energy computed as a function of two collective coordinates shown in Fig.~\ref{fig:site_dists}. In (a) we characterized the defect configuration using \(d(C_{00\frac{1}{2}})\) and in (b) using the differences of the \(d(C_{\mathrm{Na}})\) of both occupants \(u\) and \(d\) of the dumbbell.

\twocolumngrid

In Fig.~\ref{fig:site_dists}~(a) we notice that the minimum - which is the location of the dumbbell - only minimally moves away from the \((0,0,\frac{1}{2})\) site as temperature increases. Except between 900 and \SI{1000}{\kelvin}, where it seems to go through a transition and actually goes to a new minimum. This is similar to what was observed in the relaxed ground state of the dumbbell across temperature~\cite{gorfer2024structure}, which also went through a transition. There it was a discontinuity in which the orientation of the dumbbell jumped discontinuously around \SI{752}{\kelvin}, aligning itself closer to the [010] direction. The discontinuity that we notice here might be a related effect.

\section{Conclusions}\label{sec:conclusions}

In this study we performed molecular dynamics simulations of Na-feldspar with interstitial \ch{Na^+_I} and vacancy \ch{V^-_{Na}} defects. We used a previously trained neural network potential \cite{gorfer2024structure} and compared the computed \ch{Na^+} self-diffusion coefficients to the three dimensional diffusion coefficient measurement of Kasper~\cite{Kasper1975} and to the more recent measurement in the \(\perp\!\!(001)\)-direction of Wilangowski et al.~\cite{wilangowski2015_proceeding}. Our results lie inside the uncertainty interval of these experiments and also exhibit the \(\perp\!\!(001) >\, \perp\!\!(010)\) anisotropy that was measured in a more broad class of cation diffusion experiments in feldspar \cite{giletti_1974, christoffersen1983, Schaeffer2014Sodium, ElMaanaoui2016}. However, the anisotropy is small enough such that it also aligns with the findings of Behrens et al.~\cite{Behrens1990}, who found \ch{Na^+} self-diffusion in mixed Ca-Na-feldspar to be isotropic at \SI{1073.15}{\kelvin} and \SI{1173.15}{\kelvin}. In comparison, our calculations span a wider temperature range from \SI{900}{\kelvin} to \SI{1400}{\kelvin}, providing more confidence to the notion of an anisotropic diffusivity. Petrishcheva and coworkers~\cite{Petrishcheva2014} found Na-K interdiffusion in K-rich alkali feldspar to be markedly anisotropic with the slowest direction being \(\perp\!\!(10\overline{1})\), which aligns well with our slowest direction \(\perp\!\!(20\overline{1})\).

The good agreement between our model simulations and the experimental data justified a detailed investigation into the mechanism of alkali transport. As was argued in \cite{ElMaanaoui2016} and \cite{wilangowski2015_proceeding}, the interstitial defect was found to be the major contributor to \ch{Na^+} transport. We observed the two types of Na$^+$ interstitials that were previously predicted~\cite{gorfer2024structure}, a double occupancy of the alkali site forming a dumbbell type defect, and an Na$^+$ interstitial at the \((0,0,\tfrac{1}{2})\)-site, where the former is the by far dominant Na$^+$ interstitial.
For three-dimensional diffusion the contribution of Na$^+$ interstitials to the diffusion coefficient varies from 14 to 27 times larger than that of vacancies in the temperature range 1000-\SI{1400}{\kelvin}.

The mobility of this interstitial in turn is dominated by reactions going from dumbbell to dumbbell. Only rarely a transition to an \ch{Na^+_{\((00\tfrac{1}{2})\)}} defect occurs, which aligns with the fact that this state has a higher formation energy~\cite{gorfer2024structure}. At the dumbbell defect we found seven relevant migration paths and a rotation for which we calculated transition rates. Three paths (\(\mathbf{a}_1\), \(\mathbf{a}_2\) and \(\mathbf{a}_3\)) were found to dominate the overall rates and based on their arrangement we determined the slow direction \(\perp\!\!(20\overline{1})\). In light of the energetically favourable \(\mathbf{a}_2\) and \(\mathbf{a}_3\) paths, which are the modalities in which \ch{Na^+} is transported in the \(\perp\!(010)\)-direction, the anisotropy of the previous paragraph might seem surprising. The reason for the discrepancy is a markedly low correlation factor in this direction, such that the mobility of the interstitial defect is significantly higher than that of the \ch{Na^+} cations it transports. This is because the dumbbell is oriented along the \(\perp\!(010)\) direction such that the upper \ch{Na^+} can almost only move up. But after such a move it is almost certainly in a down position and can almost only move down. Actual \(\perp\!(010)\) movement therefore requires a rotational transition or some other transition that could be summarized as a migration plus rotation.

The rates - of which there are 25 in the case of the interstitial mechanism - could be readily applied to kinetic Monte Carlo simulations to improve our estimates of the correlation factors which have large uncertainties, particularly in the - for ionic conductivity measurements - important one dimensional cases.

\section{Supplementary Material}
See the supplementary material for all Arrhenius fit parameters of jump rates and diffusion coefficients, residence times, correlation factors across temperature and mean square displacement curves.

\section{Acknowledgments}
We acknowledge the financial support from the Austrian Science Fund (FWF) through Grant No.~I~4404 and through the SFB TACO, Grant No.~F-81. For open access purposes, the author has applied a CC BY public copyright license to any author accepted manuscript version arising from this submission. The computational results presented were achieved using the Vienna Scientific Cluster (VSC).

\end{document}


\onecolumngrid
\begin{center}
\textbf{\large Supplemental Materials: Mechanism and kinetics of sodium diffusion in Na-feldspar from neural network based atomistic simulations}
\end{center}
\setcounter{section}{0}
\renewcommand{\thesection}{S-\Roman{section}}
\setcounter{equation}{0}
\renewcommand{\theequation}{S-\arabic{equation}}
\setcounter{table}{0}
\renewcommand{\thetable}{S-\arabic{table}}
\setcounter{figure}{0}
\renewcommand{\thefigure}{S-\arabic{figure}}
\newcounter{SIfig}
\renewcommand{\theSIfig}{S-\arabic{SIfig}}

\section{Arrhenius Parameters}
The jump rates as well as the diffusion coefficients conform well to the Arrhenius equation
\begin{equation*}
    k = A e^{\frac{-E_a}{k_\mathrm{B}T}},
\end{equation*}
where we fit for a temperature-dependent quantity \(k\) the pre-exponential factor \(A\) and the activation energy \(E_a\). 

\begin{figure}[H]
    \centering
    \makebox[\columnwidth][c]{\includegraphics[width=0.5\columnwidth]{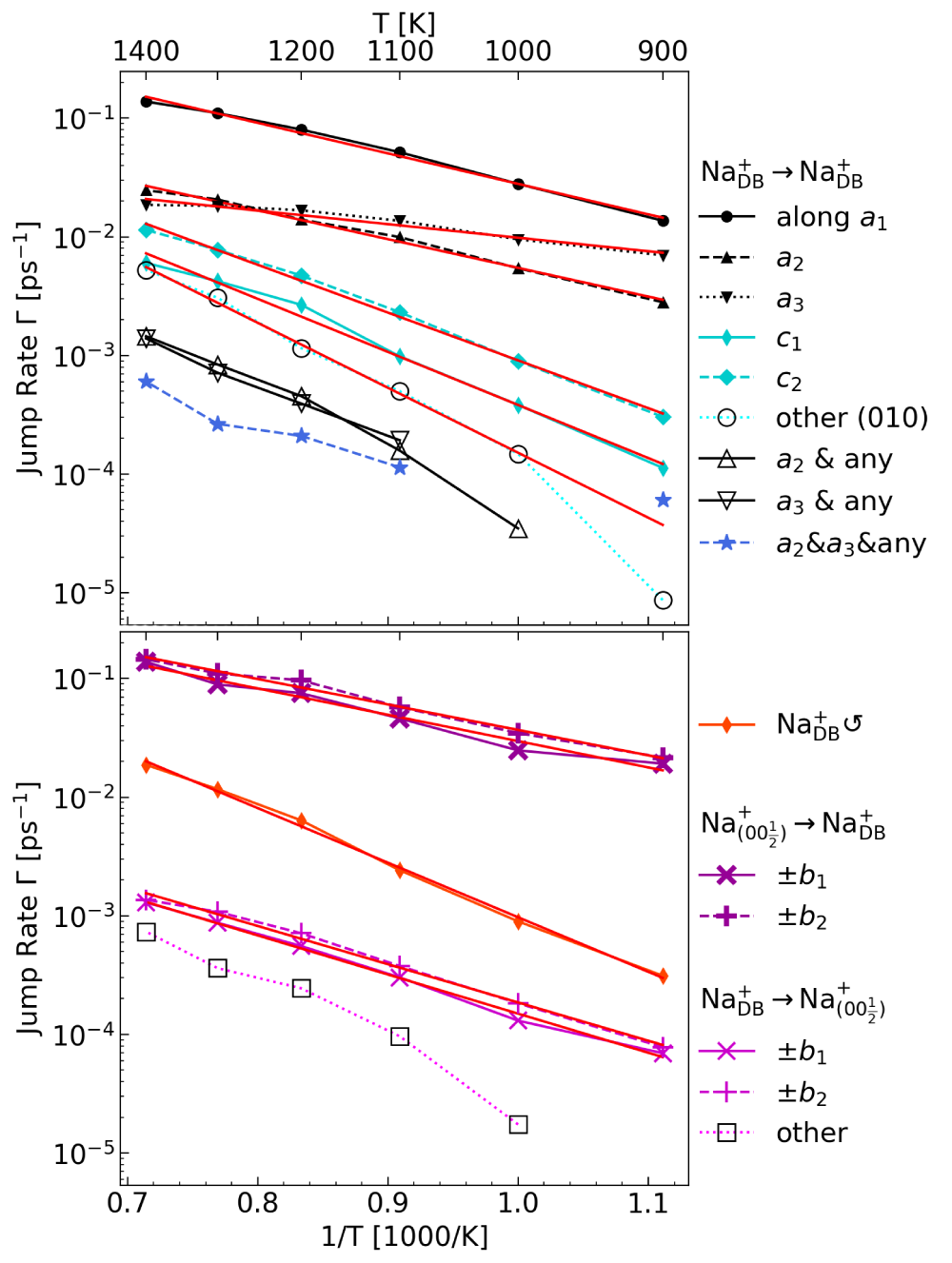}
    }
    \caption{Jump rates of the interstitial defect (as shown in Fig.~4) with the Arrhenius fits overlaid in red. The rate for “other (010)” at \SI{900}{\kelvin} was not included in its fit. The parameters are shown in Tab.~\ref{Tab:Arrhenius_hops_int}.
    }   \label{fig:hops_int_arrhenius}
\end{figure}

\begin{figure}[H]
    \centering
    \makebox[\columnwidth][c]{\includegraphics[width=0.5\columnwidth]{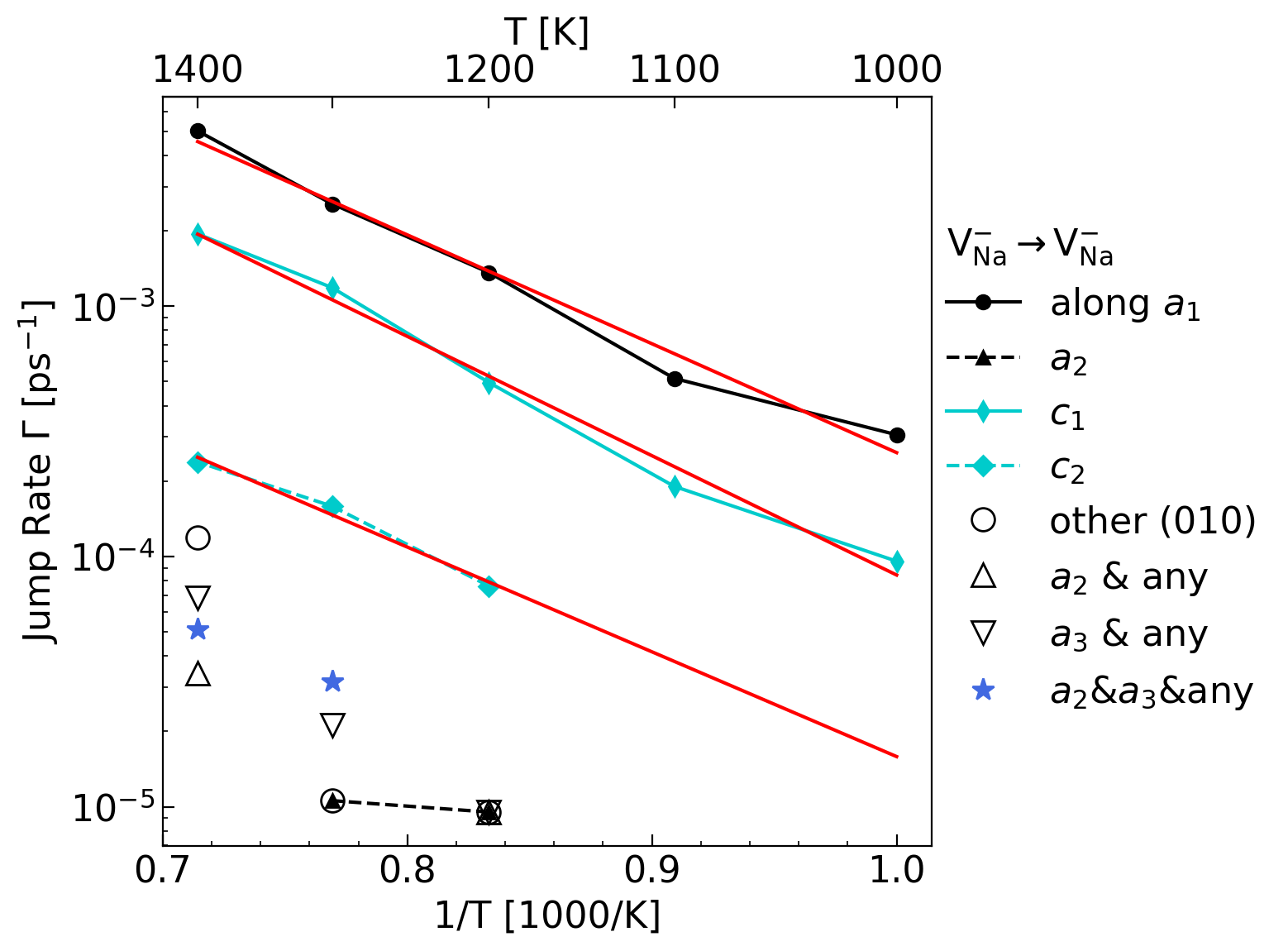}
    }
    \caption{Jump rates of the vacancy defect (as shown in Fig.~5) with the Arrhenius fits overlaid in red. The parameters are shown in Tab.~\ref{Tab:Arrhenius_hops_vac}.
    }   \label{fig:hops_vac_arrhenius}
\end{figure}

In Fig.~\ref{fig:hops_int_arrhenius} we show the fits for the jump rates of the interstitial and in Tab.~\ref{Tab:Arrhenius_hops_int} their respective parameters. The rate for “other (010)” at \SI{900}{\kelvin} was not included in the fit. The Arrhenius form fits our data well. A small deviation may exist from \SI{1200}{\kelvin} to \SI{1400}{\kelvin} as hops in \(\mathbf{a}_1\), \(\mathbf{a}_3\), \(\mathbf{c_1}\) and the rotation \(\ch{Na^+_{DB}}\circlearrowleft\) all show a minimally more gentle slope at these higher temperatures. Figure~\ref{fig:hops_vac_arrhenius} and Tab.~\ref{Tab:Arrhenius_hops_vac} show the Arrhenius fits for the hops of the vacancy. 

\begin{figure}[H]
    \centering
    \makebox[\columnwidth][c]{\includegraphics[width=0.5\columnwidth]{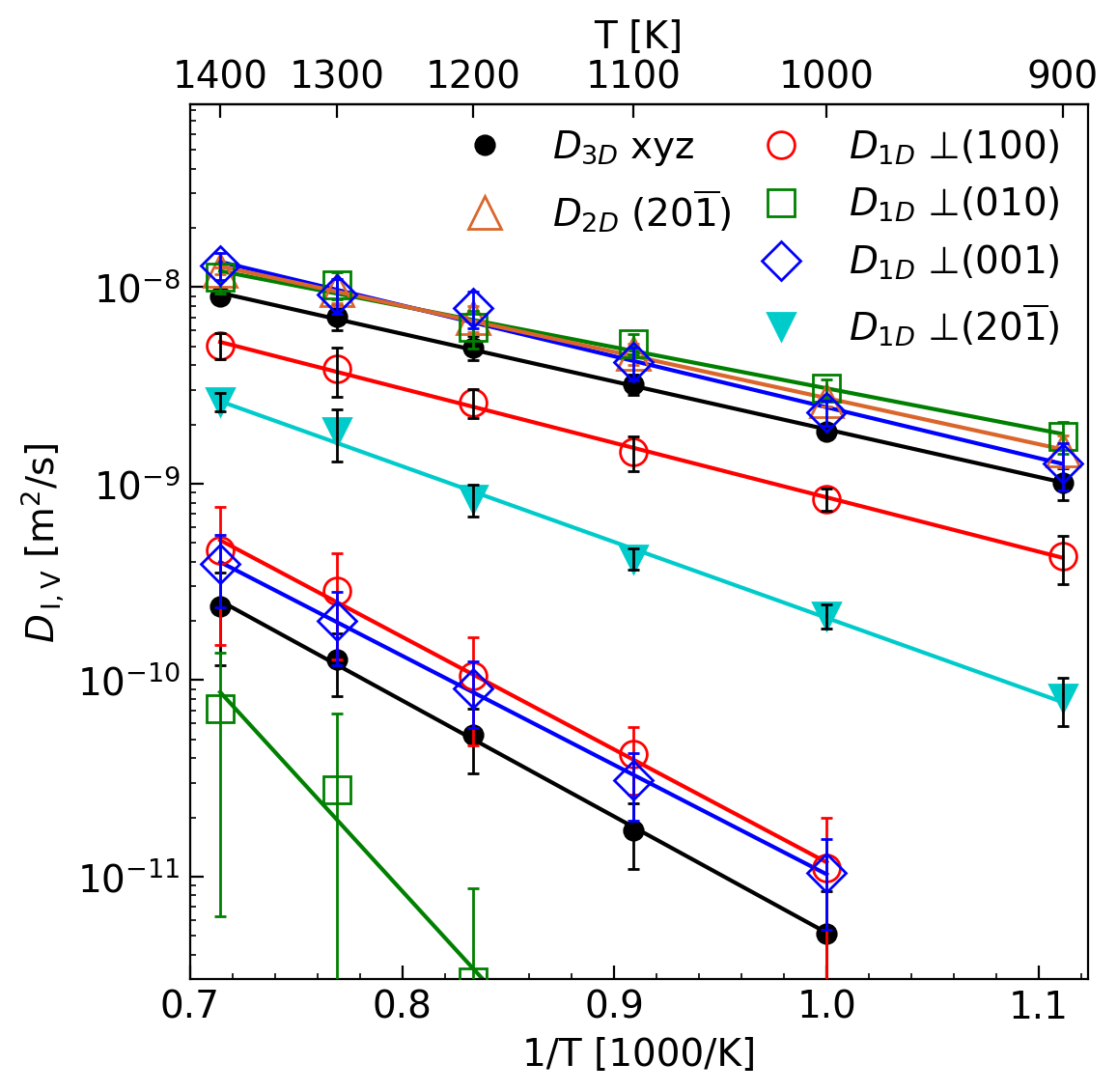}
    }
    \caption{Arrhenius fits of the Diffusion coefficients of the interstitial \(D_{\mathrm{I}}\) (upper 6) and of the vacancy \(D_{\mathrm{V}}\) (lower 4) as shown in Fig.~6. The parameters can be found in Tab.~\ref{Tab:Arrhenius_defect_diffusion}.
    }   \label{fig:dcoeff_arrhenius}
\end{figure}

In Fig.~\ref{fig:dcoeff_arrhenius} we see the Arrhenius fits for the defect-diffusion coefficents \(D_\mathrm{I}\) and \(D_\mathrm{V}\) calculated using Eq.~(5). The parameters can be found in Tab.~\ref{Tab:Arrhenius_defect_diffusion}.

In Tab.~\ref{Tab:dcoeff_arrhenius_na_diff} the respective Arrhenius parameters of the self-diffusivity per defect shown in Figure~8~(b) as well as their sum times the defect concentration calculated in~\cite{gorfer2024structure}, which is the sodium self-diffusion coefficient of Eq.~(7) shown in Figure~8~(a) are available.

\begin{table}[H]
    \centering
    \caption{Arrhenius parameters for the hops of the interstitial defect shown in Fig.~\ref{fig:hops_int_arrhenius}.}
    \begin{tabular}{c c c }
      Jump direction  & \(A\) [\si{\per\pico\second}] &  \(E_a\) [eV]  \\\toprule
      \multicolumn{3}{c}{\(\ch{Na^+_{DB}}\rightarrow\ch{Na^+_{DB}}\)}\\
      \(\mathbf{a}_1\)   & 10.37 & 0.51  \\
      \(\mathbf{a}_2\)   & 1.44  & 0.48  \\ 
      \(\mathbf{a}_3\)   & 0.14 & 0.23  \\
      \(\mathbf{c}_1\)   & 11.36 & 0.89  \\
      \(\mathbf{c}_2\)   & 9.68 & 0.80  \\
      other (010)        & 45.34 & 1.09  \\\hline     %
      \multicolumn{3}{c}{\(\ch{Na^+_{DB}}\circlearrowleft\)}\\
      \(\ch{Na^+_{DB}}\circlearrowleft\)   & 38.73 & 0.91  \\\hline
      \multicolumn{3}{c}{\(\ch{Na^+_{\((00\tfrac{1}{2})\)}}\rightarrow\ch{Na^+_{DB}}\)}\\
      \(\mathbf{b}_1\)   & 4.92 & 0.44  \\
      \(\mathbf{b}_2\)   & 5.15 & 0.43  \\\hline
      \multicolumn{3}{c}{\(\ch{Na^+_{DB}}\rightarrow\ch{Na^+_{\((00\tfrac{1}{2})\)}}\)}\\
      \(\mathbf{b}_1\)   & 0.30 & 0.65  \\
      \(\mathbf{b}_2\)   & 0.31 & 0.64  \\\hline
    \end{tabular}
    \label{Tab:Arrhenius_hops_int}
\end{table}

\begin{table}[H]
    \centering
    \caption{Arrhenius parameters for the hops of the vacancy shown in Fig.~\ref{fig:hops_vac_arrhenius}.}
    \begin{tabular}{c c c }
      Jump direction  & \(A\) [\si{\per\pico\second}] &  \(E_a\) [eV]  \\\toprule
      \multicolumn{3}{c}{\(\ch{V^-_{Na}}\rightarrow\ch{V^-_{Na}}\)}\\
      \(\mathbf{a}_1\)   & 5.80 & 0.86  \\
      \(\mathbf{c}_1\)   & 4.91 & 0.95  \\
      \(\mathbf{c}_2\)   & 0.24 & 0.83  \\\hline
    \end{tabular}
    \label{Tab:Arrhenius_hops_vac}
\end{table}

\begin{table}[H]
    \centering
    \caption{Arrhenius parameters of the defect diffusion coefficients shown in Fig.~\ref{fig:dcoeff_arrhenius}.}
    \begin{tabular}{l c c }
      Direction  & \(A\) [\SI{1e-7}{\meter\squared\per\second}] &  \(E_a\) [eV]  \\\toprule
      \multicolumn{3}{c}{Interstitial \(D_{\mathrm{I}}\)}\\
      \(D_{3D}\) xyz              & 5.09 & 0.48  \\
      \(D_{1D} \perp(100)\)       & 4.91 & 0.95  \\
      \(D_{1D} \perp(010)\)       & 3.75 & 0.41 \\
      \(D_{1D} \perp(001)\)       & 9.54 & 0.51 \\
      \(D_{2D} ~(20\overline{1})\)& 6.07 & 0.47 \\
      \(D_{1D} \perp(20\overline{1})\)& 14.97 & 0.77 \\\hline
      \multicolumn{3}{c}{Vacancy \(D_{\mathrm{V}}\)}\\
      \(D_{3D}\) xyz & 42.23 & 1.17\\
      \(D_{1D} \perp(100)\) & 64.63 & 1.14\\
      \(D_{1D} \perp(010)\) & \num{24e+4} & 2.34\\
      \(D_{1D} \perp(001)\) & 37.39 & 1.10\\\hline
    \end{tabular}
    \label{Tab:Arrhenius_defect_diffusion}
\end{table}

\begin{table}[H]
    \centering
    \caption{Arrhenius parameters for sodium self-diffusion (Fig.~8), separately for the interstitial and for the vacancy contribution (see Eq.~3) and their sum, the total \(D_{\ch{Na}}\) which also includes the concentration (see Eq.~7).}
    \begin{tabular}{l c c }
      Direction  & \(A\) [\SI{1e-7}{\meter\squared\per\second}] &  \(E_a\) [eV]  \\\toprule
      \multicolumn{3}{c}{Interstitial \(D^{(\mathrm{I})}_{\ch{Na}}\)}\\
        \(D_{3D}\) xyz               & 4.20 & 0.49 \\
        \(D_{1D} \perp(100)\)        &5.32 & 0.56 \\
        \(D_{1D} \perp(010)\)        &2.59 & 0.41 \\
        \(D_{1D} \perp(001)\)        &8.11 & 0.51 \\
        \(D_{2D} ~(20\overline{1})\) & 4.83 & 0.47 \\
        \(D_{1D} \perp(20\overline{1})\) & 11.80 & 0.74 \\\hline
        \multicolumn{3}{c}{Vacancy \(D^{(\mathrm{V})}_{\ch{Na}}\)}\\
        \(D_{3D}\) xyz                   & 143.29 & 1.37 \\
        \(D_{1D} \perp(100)\)            & 35.63 & 1.18 \\
        \(D_{1D} \perp(010)\)            & 548.79 & 1.60 \\
        \(D_{1D} \perp(001)\)            & 198.24 & 1.35 \\
        \(D_{2D} ~(20\overline{1})\)     & 377.22 & 1.48 \\
        \(D_{1D} \perp(20\overline{1})\) & 28.81 & 1.20 \\\hline
        \multicolumn{3}{c}{\(D_{\ch{Na}}\)}\\
        \(D_{3D}\) xyz                   & 13.29 & 1.53 \\
        \(D_{1D} \perp(100)\)            & 17.61 & 1.60 \\
        \(D_{1D} \perp(010)\)            & 7.97 &  1.45 \\
        \(D_{1D} \perp(001)\)            & 25.67 & 1.55 \\
        \(D_{2D} ~(20\overline{1})\)     & 15.12 & 1.51 \\
        \(D_{1D} \perp(20\overline{1})\) & 40.96 & 1.79 \\\hline
    \end{tabular}
    \label{Tab:dcoeff_arrhenius_na_diff}
\end{table}

\newpage
\section{Residence times}

\begin{figure}[h!]
    \centering
    \makebox[\columnwidth][c]{\includegraphics[width=0.5\columnwidth]{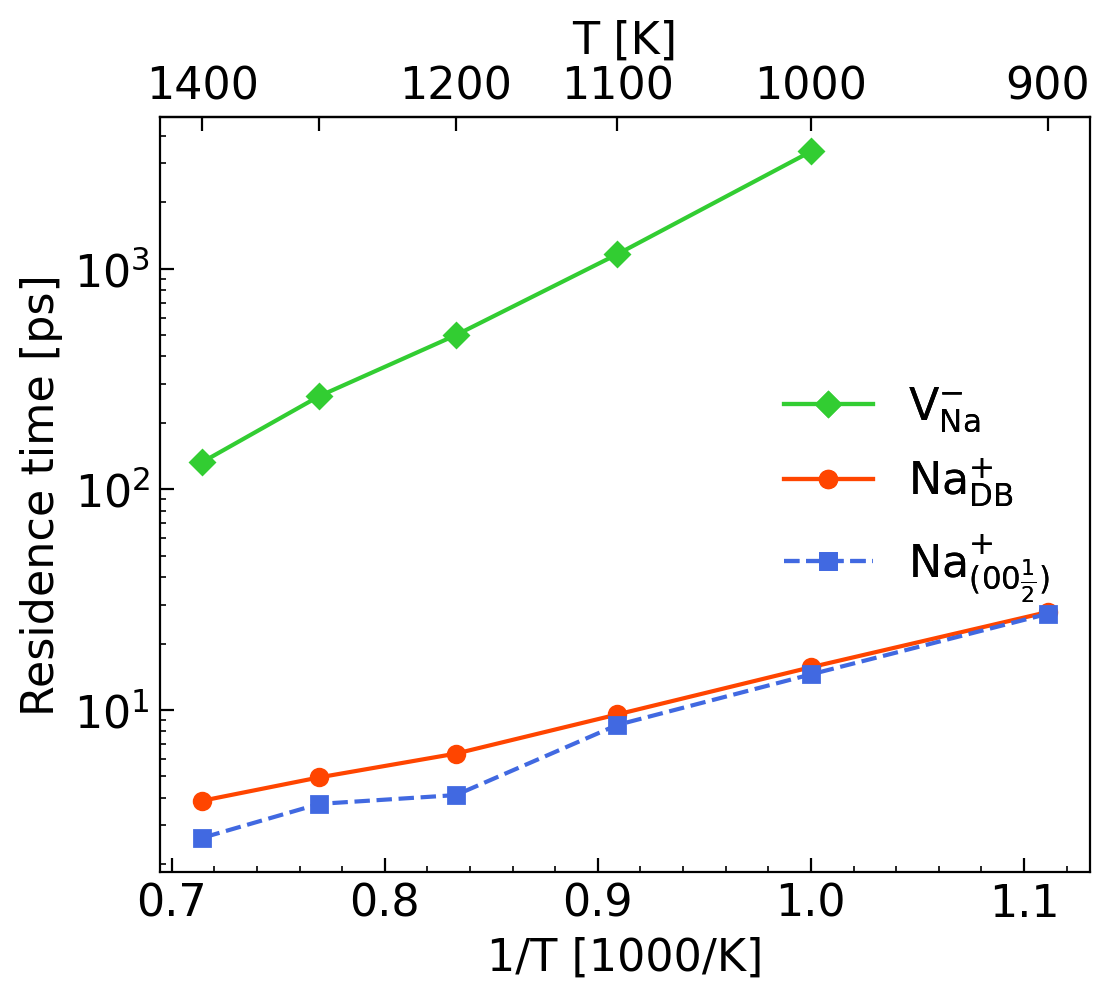
    }}
    \caption{Residence time of the different defects.
    }   \label{fig:lifetime}
\end{figure}

In Fig.~\ref{fig:lifetime} the residence times of all defects are plotted. Since exit rates for \(\ch{Na^+_{\((00\tfrac{1}{2})\)}}\rightarrow\ch{Na^+_{DB}}\) are similar in size to the rates of \(\ch{Na^+_{DB}}\rightarrow\ch{Na^+_{DB}}\) along \(\mathbf{a}_1\), \(\ch{Na^+_{DB}}\) and \(\ch{Na^+_{\((00\tfrac{1}{2})\)}}\) fall quite close to each other and from 900 to about \SI{1100}{\kelvin}, \(\ch{Na^+_{DB}}\) and \(\ch{Na^+_{\((00\tfrac{1}{2})\)}}\) seem to almost agree completely.

\section{Correlation factors}

\begin{figure}[h]
    \centering
    \makebox[\columnwidth][c]{\includegraphics[width=0.5\columnwidth]{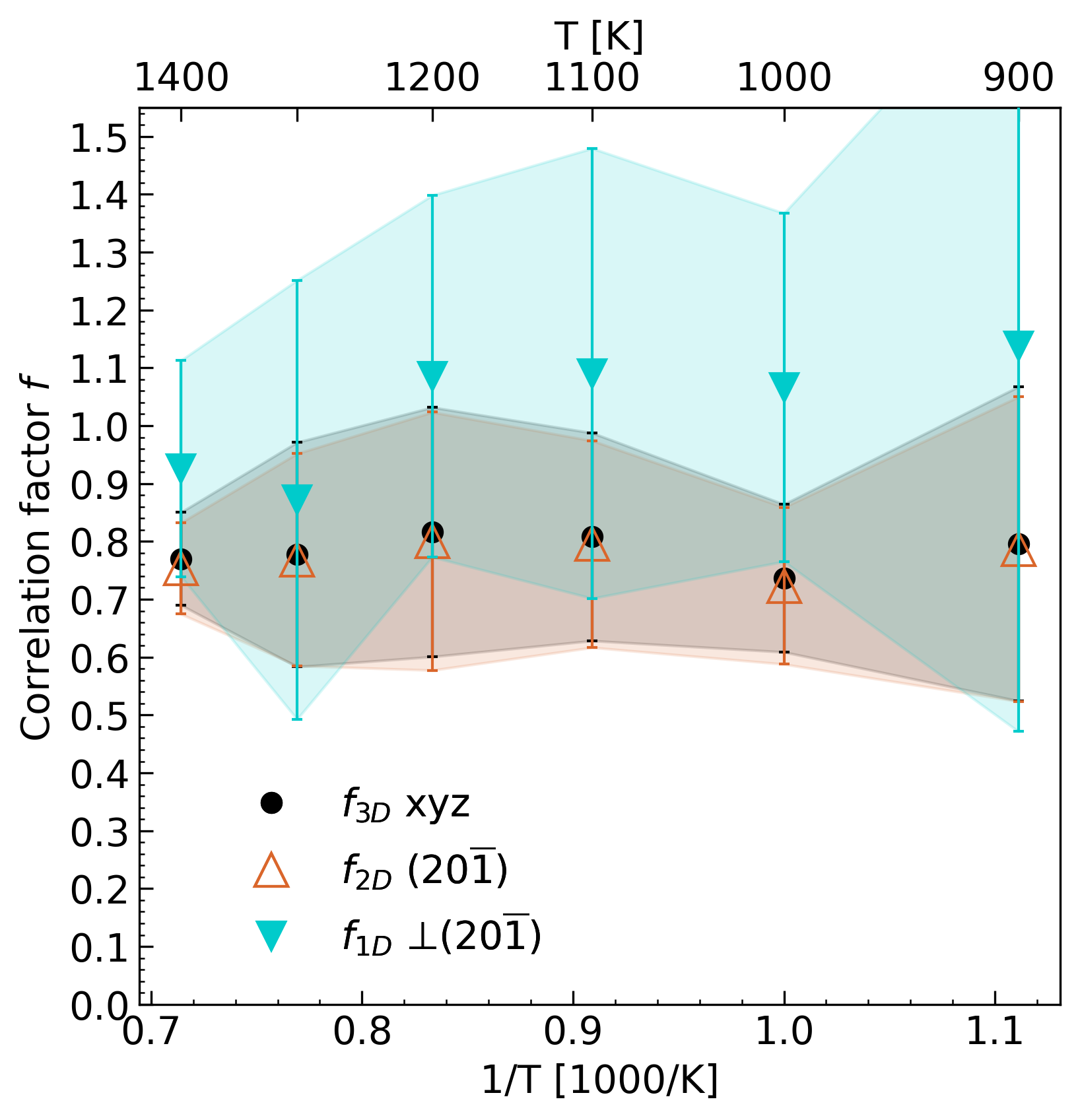
    }}
    \caption{Correlation factors of interstitial diffusion calculated in Eq.~(6) for three directionalities.
    }   \label{fig:CorrelationFactor1}
\end{figure}

\begin{figure}[h]
    \centering
    \makebox[\columnwidth][c]{\includegraphics[width=0.5\columnwidth]{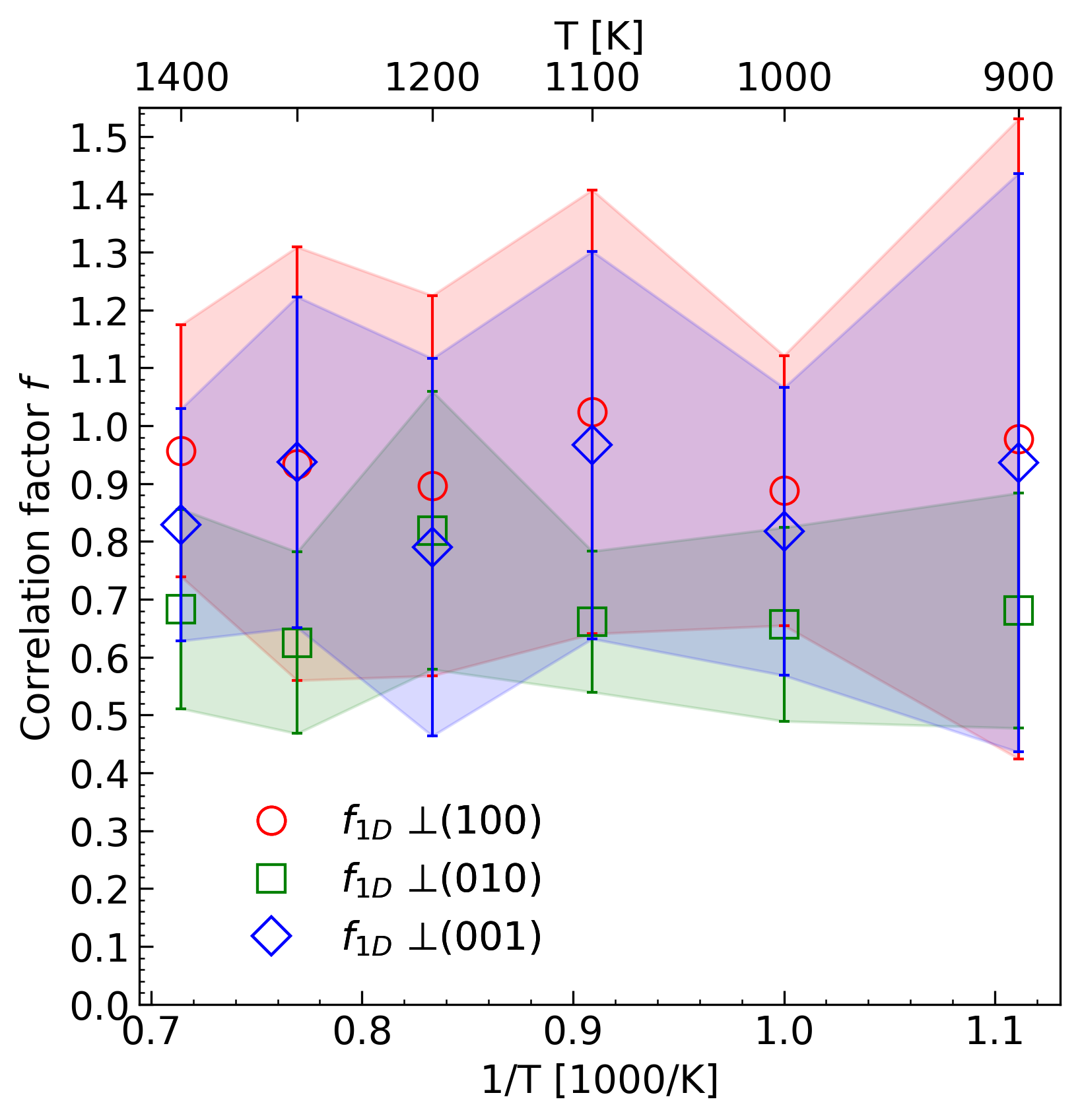
    }}
    \caption{Correlation factors of interstitial diffusion calculated in Eq.~(6) for the other three directionalities.
    }   \label{fig:CorrelationFactor2}
\end{figure}

Figures \ref{fig:CorrelationFactor1} and \ref{fig:CorrelationFactor2} show the temperature dependence of the correlation factors of interstitial diffusion calculated using Eq.~(6). The average over temperature shown in Tab.~I is inside the uncertainty range of these values.

\section{Mean Square displacements}
\begin{figure}[h]
    \centering
    \makebox[\columnwidth][c]{\includegraphics[width=1.0\columnwidth]{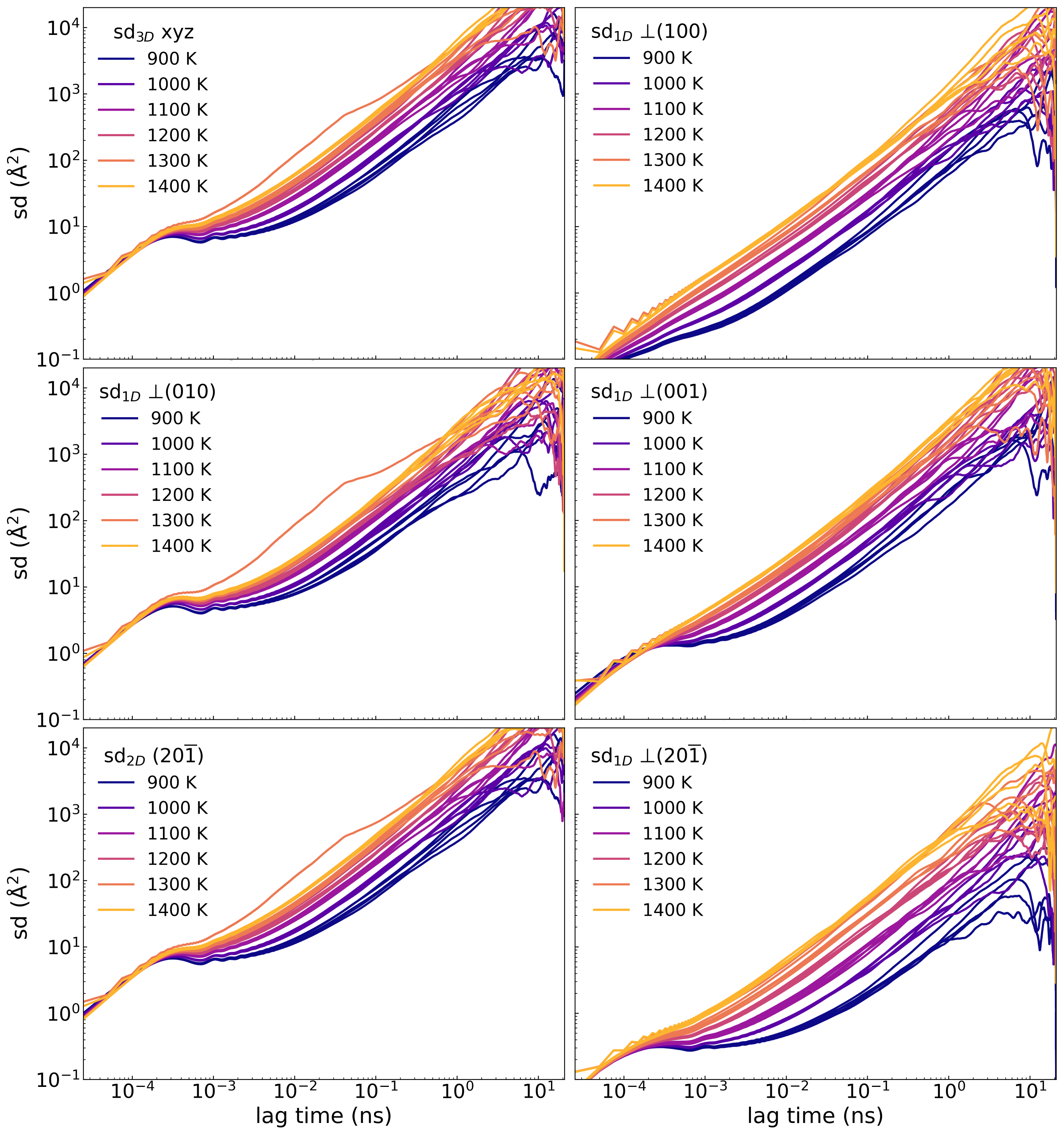}
    }
    \caption{Mean square displacements of the interstitial defect (see Eq.~4) in a log-log view for the respective diffusion directionalities. All five independent trajectories are shown. One of these, at \SI{1300}{\kelvin}, can be considered an outlier. It is the reason why the measurement uncertainty of interstitial diffusion coefficients at \SI{1300}{\kelvin} is markedly higher than at \SI{1200}{\kelvin} or \SI{1100}{\kelvin}.
    }   \label{fig:msd_ints}
\end{figure}

\begin{figure}[t]
    \centering
    \makebox[\columnwidth][c]{\includegraphics[width=1.0\columnwidth]{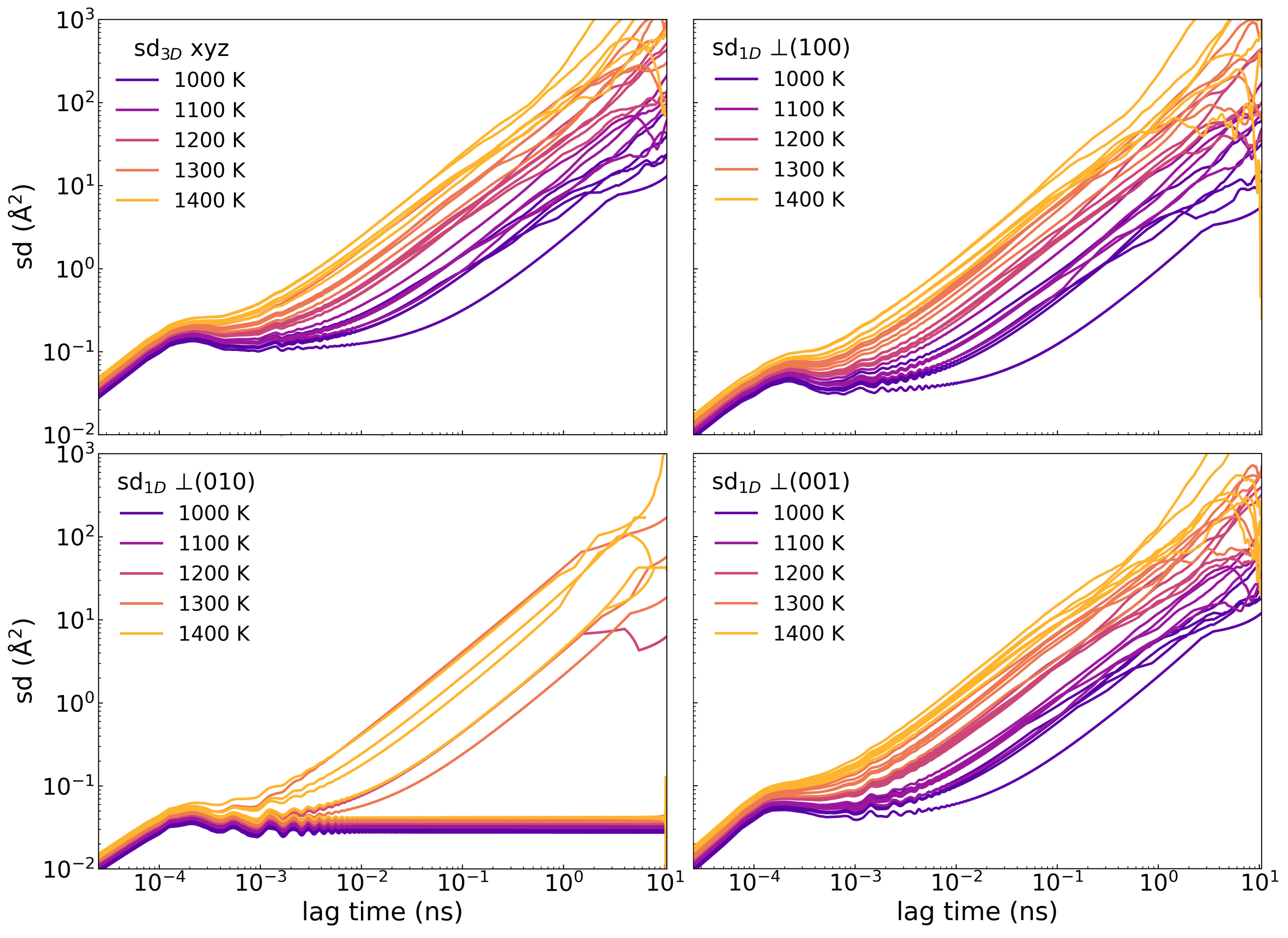}
    }
    \caption{Mean square displacements of the vacancy defect (see Eq.~4) in a log-log view for the respective diffusion directionalities. All five independent trajectories are shown. For the \(\perp(010)\) direction only some trajectories show any movement at all. That is 4 out of 5 at \SI{1400}{\kelvin}, 3 at \SI{1300}{\kelvin} and only 1 at \SI{1200}{\kelvin}. Below \SI{1200}{\kelvin} the vacancy stayed in the (010) plane.
    }   \label{fig:msd_vacs}
\end{figure}

The mean square displacements of the defects shown in Eq.~(4) is plotted in Figures~\ref{fig:msd_ints} and \ref{fig:msd_vacs}. All five trajectories are shown for each temperature. In the case of the interstitial there is an outlier at \SI{1300}{\kelvin} which is the origin behind the high uncertainty at \SI{1300}{\kelvin} in the diffusion coefficients. A peculiarity for the vacancy mean square displacements is the direction \(\perp(010)\) where only a few trajectories show any movement past the ballistic regime. That is 4 out of 5 at \SI{1400}{\kelvin}, 3 out of 5 at \SI{1300}{\kelvin} and only a single one at \SI{1200}{\kelvin}. Below \SI{1200}{\kelvin} the vacancy stayed in the (010) plane.

